\documentclass[prc,showpacs,floatfix]{revtex4}
\usepackage{bm}
\usepackage{graphicx}
\newcommand{\slt}{\!\!\!/}
\newcommand{\sld}{\!\!/}
\newcommand{\beq}{\begin{equation}}
\newcommand{\eeq}{\end{equation}}
\newcommand{\beqa}{\begin{eqnarray}}
\newcommand{\eeqa}{\end{eqnarray}}

\begin{document}

\title{
\hfill{\small {\bf MKPH-T-04-06}}\\
{\bf Interaction effects in ${\bm K^+}$ photoproduction on the deuteron}}

\author{Agus Salam\footnote{Present address: Departemen Fisika, FMIPA, 
Universitas Indonesia, Depok 16424, Indonesia.} and Hartmuth
Arenh\"ovel} 
\affiliation{Institut f\"{u}r Kernphysik, Johannes
Gutenberg-Universit\"{a}t, D-55099 Mainz, Germany} 
\date{\today}
\begin{abstract} 
Kaon photoproduction on the deuteron is studied with respect to a 
specific two-body contribution, namely a pion mediated 
production process, besides other final state
interaction contributions from kaon-nucleon and hyperon-nucleon
scattering. In this process, a pion is first photoproduced on one
nucleon and then interacts with the spectator nucleon in a
strangeness exchange reaction leading to a kaon and a hyperon. A
sizeable effect from this pion mediated contribution is found,
considerably larger than the previously studied hyperon-nucleon
rescattering, whereas kaon-nucleon rescattering is much less important. 
Besides total and semi-inclusive differential cross sections, tensor
target asymmetries are studied with respect to the influence of
such interaction effects. 
\end{abstract}
\maketitle

%%%%%%%%%%%%%%%%%%%%%%%%%%%%%%%%%%%%%%%%%%%%%%%%%%%%%%%%%%%%%%%%%%%%%%%%%%%%%%%
\section{Introduction}
\label{sec1}

The study of kaon photoproduction has drawn attention for more than
three decades since the work of Thom~\cite{Tho66}, who analyzed the
reaction $\gamma$p$\rightarrow$K$^+\Lambda$ by using Feynman diagrams
for the Born terms and partial wave amplitudes for the
resonances. Adelseck {\it et al.}~\cite{AdB85} evaluated the resonance
terms using diagrammatic 
techniques, in order to ensure the relativistic invariance of the
operator. The advent of a new generation of high duty-factor
accelerators of sufficiently high energy such as MAMI in Mainz,
ELSA in Bonn, or CEBAF in Newport News, has triggered several new
analyses. David {\it et al.}~\cite{DaF96} have analyzed the
strangeness production by including in addition spin-5/2
resonances and off-shell effects of the resonance vertices. Mart and
Bennhold~\cite{Mar96,BeM96} have included an overall hadronic form
factor in the production operator. This work was refined in
\cite{Mar00,LeM01} by allowing different hadronic form factors at the
various vertices in conjunction with the recipe of
Haberzettl~\cite{HaB98} in order to ensure gauge 
invariance. Other models were used in~\cite{HaC01} investigating
pseudovector coupling and in~\cite{JaR02} studying the role of hyperon
resonances in kaon photoproduction off the nucleon. 

The channels mostly investigated in kaon photoproduction are the
proton channels, $\gamma p\rightarrow K^+\Lambda$ and
$\gamma p\rightarrow K^+\Sigma^0$, in view of a
relatively large number of experimental data for these channels
\cite{Boc94,Tra98}. Because the neutron has a short lifetime, free neutron
targets are not available for the study of the neutron channels, and
thus one uses light nuclei like 
deuterium or $^3$He as effective neutron targets. The deuteron is
particularly suited because of its small binding energy and its simple
structure. With the purpose to extract the
elementary cross section on a neutron target, Li {\it et
al.}~\cite{LiW92} have calculated the reactions 
$\gamma d \rightarrow K ^0\Lambda p$,
$\gamma d \rightarrow K ^0\Sigma^0 p$, and
$\gamma d \rightarrow K ^+\Sigma^- p$ in the impulse approximation (IA)
only. They concluded that the deuteron can be used to study K$^0$ and
K$^+$ photoproduction from the neutron. The study of the
hyperon-nucleon interaction is another important aspect of
of kaon photoproduction on the deuteron. Several investigations of
this question exist already. Renard and Renard \cite{ReR67a,ReR67b}
have derived 
the formalism and studied the $\Lambda n$ interaction in kaon
photoproduction off the deuteron. Adelseck and Wright \cite{AdW89}
have examined the $\Lambda n$ final state interaction in kaon
photoproduction from the deuteron via a distorted wave formalism by
using a simple $\Lambda n$ potential. With the intention of investigating
the hyperon-nucleon interaction, in a recent paper Yamamura {\it et
al.}~\cite{YaM99} have calculated the hyperon-nucleon final state
interaction for the K$^+$ channels by using the more realistic Nijmegen
$YN$ potential from~\cite{MaR89,RiS99}. They found sizeable effects in both, 
exclusive as well as inclusive cross sections from the $YN$ interaction, 
in particular a cusplike structure near the production threshold of the
$\Sigma$ channels, and concluded that precise data would allow to study 
the $YN$ interaction in greater detail. Another recent calculation is
from Kerbikov~\cite{Ker01} who also investigated the hyperon-nucleon
final state interaction.

Thus up to now, of the various interactions in the final 
three-particle state of kaon
photoproduction on the deuteron, most of the calculations have considered 
only the hyperon-nucleon final state interaction ($YN$-FSI)
quantitatively. With respect to the other two possible interactions in
the kaon-hyperon and kaon-nucleon two-body subsystems, the former is
usually assumed to be already included in the elementary production
amplitude whereas the latter has been considered as 
negligible. In the present paper, our first point of interest is the
quantitative study of this kaon-nucleon final state interaction ($KN$-FSI)
by including the kaon-nucleon scattering matrix into the 
photoproduction amplitude. Our second point refers to the inclusion of 
another competing two-body process which might give in
addition an important contribution and which has been neglected
hitherto. It refers to the pion mediated kaon production process, 
denoted by ``$\pi\rightarrow K$'', in which the absorbed photon produces 
first on one nucleon a pion which then interacts 
with the specator nucleon via a strangeness exchange reaction leading
to a kaon and  a hyperon. Although on first sight this process,
being a two-step reaction, is expected to be suppressed, it could give
a sizeable contribution in view of the fact that the pion
photopoduction cross section is still relatively strong in the region
of kaon photoproduction. For both
hadronic reactions, $KN$ scattering and $\pi \rightarrow K$, 
separable potentials are used. After completion of this work, a very 
recent study of two-body contributions to the photoproduction 
operator was published by Maxwell~\cite{Max04} using a diagrammatic 
approach where the pion mediated process is included in lowest order. 
Our approach differs with respect to the one-body photoproduction 
operator on the nucleon and with respect to a complete inclusion of 
the various two-body reactions in the two-body subsystems. 
In particular, final state correlations like hyperon-nucleon and 
kaon-nucleon rescattering were not included in~\cite{Max04}.

In Sect.~\ref{sec2}, we
briefly review the formal aspects of the various two-body elementary
reactions which we include in our treatment of kaon
photoproduction on the deuteron. In Sect.~\ref{sec3}, the formalism for
calculating the transition matrix and cross section for kaon
photoproduction on the deuteron with inclusion of final state
interactions and the $\pi\rightarrow K$ process 
is given. The results are presented in Sect.~\ref{sec4} and we
close with some conclusions and an outlook in Sect.~\ref{sec5}.
Throughout the paper we use natural units $\hbar = c = 1$.

%%%%%%%%%%%%%%%%%%%%%%%%%%%%%%%%%%%%%%%%%%%%%%%%%%%%%%%%%%%%%%%%%%%%%%%%%%%%%%%
\section{Elementary reactions}
\label{sec2}

Kaon photoproduction on the deuteron is governed by basic 
two-body processes, namely meson photoproduction on a nucleon and 
hadronic two-body scattering reactions. In this section we will collect 
the necessary ingredients for the various processes which we have included 
in the present theoretical description of kaon photoproduction on the 
deuteron. 

The general form of these elementary two-body reactions is
\begin{eqnarray}
{A}(p_{A}) + {B}(p_{B}) 
&\rightarrow& 
{C}(p_{C}) + {D}(p_{D})\,,
\label{eq-abcd}
\end{eqnarray}
where $p_{i}=(E_{i},\,\vec p_{i})$ denotes the 4-momentum of particle
``$i$'' with $i\in\{A, B, C, D\}$. Particles $A$ and $B$ stand for a photon 
and a nucleon in photoproduction, a meson and a baryon in the case
of kaon-nucleon scattering and the $\pi N\rightarrow K Y$ process, or a pair of
baryons like in hyperon-nucleon scattering. Corresponding assignments stand
for the final particles $C$ and $D$. 
   
In order to compare the theoretical predictions for the various elementary 
reactions with experimental data one has to evaluate the corresponding 
cross sections. Following the conventions of Bjorken and Drell~\cite{BjD64} 
the general form for the differential cross section of a two-particle 
reaction in the center of mass system is given by
\begin{eqnarray}
\frac{d\sigma}{d\Omega_{C}} 
&=&
\frac{1}{(2\pi W)^{2}}\,
\frac{p_{C}\,F}{p_{A}\,s}
\sum_{\mu_{D}\mu_{C}\mu_{B}\mu_{A}}
\left \vert 
{\cal M}_{\mu_{D}\mu_{C}\mu_{B}\mu_{A}}
(\vec p_{D},\vec p_{C},\vec p_{B},\vec p_{A}) 
\right \vert^{2}
\label{eq-abcd-dsigma/domega}
\end{eqnarray}
with ${\cal M}_{\mu_{D}\mu_{C}\mu_{B}\mu_{A}}$ as reaction
matrix, $\mu_{i}$ denoting the spin projection of particle
``$i$'' on some quantization axis, and
\begin{eqnarray}
F &=& 
\frac{E_{A}E_{B}E_{C}E_{D}}
{F_{A}F_{B}F_{C}F_{D}}\,,
\label{eq-abcd-F}
\end{eqnarray}
where $F_i$ is a factor arising from the covariant normalization 
of the states and its form depends on whether the particle is a boson 
($F_i=2E_{i}$) or a fermion ($F_i=E_{i}/m_{i}$), where $E_{i}$ and $m_{i}$ 
are its energy and mass, respectively. The factor $s=(2s_{A}+1)
(2s_{B}+1)$ takes into account the averaging over the 
initial spin states, where $s_{A}$ and $s_{B}$ denote the spins of the
incoming particles $A$ and B, respectively. If $A$ stands for a photon then 
$s_{A}=1/2$. Note that $p_{C}$ means $|\vec
p_{C}|$. All momenta are functions of the invariant mass of the 
two-body system $W$, i.e. $p_i=p_i(W)$. 

For the scattering processes, it is more convenient to use 
non-covariant normalization of the states and to switch to a coupled 
spin representation replacing the ${\cal M}$-matrix by the 
${\cal T}$-matrix via
\begin{eqnarray}
{\cal M}^{fi}_{\mu_{D}\mu_{C}\mu_{B}\mu_{A}}
(\vec p_{D},\vec p_{C},\vec p_{B},\vec p_{A})
&=&
(2\pi)^{3} \sqrt{F_{A}F_{B}F_{C}F_{D}}
\sum_{S^{\prime} \mu^{\prime}_{S^{\prime}} S \mu_{S}} 
C^{s_{C}s_{D} S^{\prime}}_{\mu_{C}\mu_{D}\mu^{\prime}_{S^{\prime}}}\,
C^{s_{A}s_{B} S}_{\mu_{A}\mu_{B}\mu_{S}}\,
{\cal T}^{fi}_{S^{\prime} \mu^{\prime}_{S^{\prime}} S \mu_{S}}
(\vec p_{D},\vec p_{C},\vec p_{B},\vec p_{A})\,,
\label{eq-abcd-M-T-matrix}
\end{eqnarray}
with $C^{s_{C}s_{D}
S^{\prime}}_{\mu_{C}\mu_{D}\mu^{\prime}_{S^{\prime}}}$ as approriate
Clebsch-Gordan coefficient. 
As next step we introduce a partial wave representation of the
${\cal T}$-matrix which reads with $\vec p^{\,\prime}$ and $\vec p$ 
as the final and initial relative momenta, respectively, 
\begin{eqnarray}
{\cal T}^{fi}_{S^{\prime} \mu^{\prime}_{S^{\prime}} S \mu_{S}}
(W,\vec p^{\,\prime}, \vec p \,) &=&
\sum_{\ell^{\prime} \ell J}
X^{\ell^{\prime} \ell J}_{S^{\prime} \mu^{\prime}_{S^{\prime}} S \mu_{S}}
(\hat p^{\prime}, \hat p)\,
T ^{\ell^{\prime} \ell J}_{fi}(W,p^{\prime}, p)\,,
\label{eq-abcd-T-matrix-partial-wave}
\end{eqnarray}
where we have introduced
\begin{eqnarray}
X^{\ell^{\prime} \ell J}_{S^{\prime} \mu^{\prime}_{S^{\prime}} S \mu_{S}}
(\hat p^{\prime}, \hat p)
&=&
\sum_{\mu^{\prime}_{\ell^{\prime}} \mu_{\ell} \mu_{J}}
Y_{\ell^{\prime} \mu^{\prime}_{\ell^{\prime}}}(\hat p^{\prime})\,
Y_{\ell \mu_{\ell}}^{\ast}(\hat p)\,
C^{\ell^{\prime} S^{\prime} J}
_{\mu^{\prime}_{\ell^{\prime}} \mu^{\prime}_{S^{\prime}} \mu_{J}}\,
C^{\ell S J}_{\mu_{\ell} \mu_{S} \mu_{J}}\,.
\label{eq-abcd-X-function}
\end{eqnarray}
Here $\ell$ and $J$ denote the orbital and total angular momenta 
of the system, respectively, $Y_{\ell \mu}(\hat p)$ a spherical
harmonics, and $\hat p = (\theta_{\vec p},\phi_{\vec p})$. 
The partial wave ${\cal T}$-matrix is obtained as solution of the 
Lippmann-Schwinger equation
\begin{eqnarray}
{\cal T}^{\ell^{\prime}\ell J}_{fi}(W,p^{\prime},p) &=&
V^{\ell^{\prime}\ell J}_{fi}(p^{\prime},p) 
+ \sum_{n \ell^{\prime\prime}} 2m_{n} \int_{0}^{\infty} 
dp^{\prime\prime}_{n}~ (p^{\prime\prime}_{n})^{2}\,
\frac{V^{\ell^{\prime}\ell^{\prime\prime} J}_{fn}
(p^{\prime},p^{\prime\prime}_{n})\, 
T ^{\ell^{\prime\prime}\ell J}_{ni}
(W,p^{\prime\prime}_{n},p)}
{q_{n}^{2}-(p^{\prime\prime}_{n})^{2}+i\varepsilon}\,,
\label{eq-abcd-Lippmann-Schwinger-q}
\end{eqnarray}
where ``$n$'' labels possible intermediate two-particle configurations 
with total mass $M_{n}$, reduced mass $m_{n}$, and with 
relative momentum 
\begin{eqnarray}
q_{n} &=& \sqrt{2m_{n}\left(W-M_{n}\right)}
\label{eq-abcd-p0}
\end{eqnarray}
in the c.m.\ system. 
Now we will briefly review the different elementary processes in some detail.

%%%%%%%%%%%%%%%%%%%%%%%%%%%%%%%%%%%%%%%%%%%%%%%%%%%%%%%%%%%%%%%%%%%%%%%%%%%%%%%
\subsection{Kaon photoproduction on the nucleon}

\begin{figure}[htbp]
\includegraphics[width=.6\textwidth]{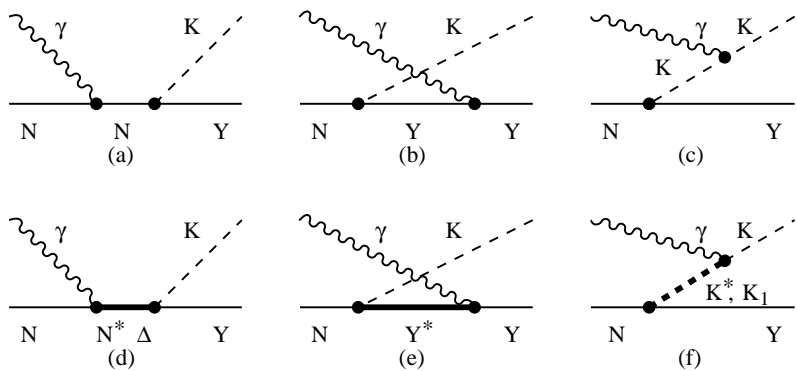}
\caption{Elementary diagrams of kaon photoproduction on the
nucleon. Born terms (a) - (c): nucleon, hyperon and kaon poles, respectively; 
resonance terms (d) - (f): nucleon, hyperon, and kaon resonance poles,
respectively.}
\label{fig-gnky-irreducible}
\end{figure}

In the simplest approach to kaon photoproduction on the nucleon, one
approximates the production amplitude ${\cal M}$ with the tree-level 
diagrams shown in Fig.~\ref{fig-gnky-irreducible}.  
In principle these diagrams serve as driving terms in a system of coupled 
equations in which hadronic rescattering is included which is important
in order to ensure unitarity. However, in the present work we use the 
simpler model of Lee et al.~\cite{LeM01}, called isobar model, in
which all diagrams of  
Fig.~\ref{fig-gnky-irreducible} are taken into account except the 
Y$^{*}$-pole diagram (e). In the Born terms pseudoscalar coupling is used 
for the hadronic meson-baryon vertices. As resonances are included
for the N$^{*}$-pole (diagram (d)) S$_{11}$(1650),
P$_{11}$(1710), S$_{31}$(1900), P$_{31}$(1910), and P$_{13}$(1720) and
for the K$^*$-pole (diagram (f)) K$^*$(892) and K$_{1}$(1270). Separate
hadronic form factors for each vertex were used which, however,
destroys gauge invariance. In order to restore gauge
invariance, a recipe from Haberzettl~\cite{HaB98} was utilized. 

The photoproduction amplitude ${\cal M}^{K\gamma N}$ is parametrized usually
in terms of four invariant operators $\Gamma^{i}_{\lambda}$, 
accompanied by invariant amplitudes $A_{i}^{K\gamma N}$ which are
functions of the Mandelstam variables only. Thus the amplitude has the
form 
\begin{eqnarray}
{\cal M}^{K\gamma N}_{\mu_{Y}\mu_{N}\lambda}
&=& 
\bar{u}_{\mu_{Y}}
\Big(\sum_{i=1}^{4} A_{i}^{K\gamma N} \Gamma^{i}_{\lambda}\Big)
u_{\mu_{N}}\,,
\label{eq-gnky-M-matrix-Dirac-spinor}
\end{eqnarray}
where we have suppressed the dependence on the
kinematical variables. The hyperon and nucleon Dirac spinors are
denoted by $u_{\mu_{Y}}$ and $u_{\mu_{N}}$, respectively. 
The invariant Dirac operators $\Gamma_{i}$ are
gauge invariant Lorentz pseudoscalars and given in terms of the usual
$\gamma$-matrices, the photon momentum $k$, its polarization vector
$\epsilon_{\lambda}$, where $\lambda$ labels the polarization states,
the meson momentum $q$ and $P=(p'+p)/2$, where $p$ and $p'$ denote
initial and final baryon momenta, respectively~\cite{Don72},
\begin{eqnarray}
\Gamma^{1}_{\lambda} 
& = & {\textstyle \frac{1}{2}} \gamma_{5} 
\left({\epsilon\sld}\!_{\lambda} k \slt 
- k \slt {\epsilon\sld}\!_{\lambda} \right)\,, \\
\Gamma^{2}_{\lambda} 
& = & \gamma_{5} \left[ (2q-k) \cdot \epsilon_{\lambda} P \cdot k 
- (2q-k) \cdot k P \cdot \epsilon_{\lambda} \right]\,, \\
\Gamma^{3}_{\lambda} 
& = & \gamma_{5} \left( q\cdot k {\epsilon\sld}\!_{\lambda}
- q \cdot \epsilon_{\lambda} k \slt \right)\,, \\
\Gamma^{4}_{\lambda} 
& = & i \epsilon_{\mu \nu \rho \sigma} \gamma^{\mu} q^{\nu}
\epsilon^{\rho}_{\lambda} k^{\sigma}\,.
\label{eq-gnky-Gamma-matrix}
\end{eqnarray}
The contributions of the various diagrams in
Fig.~\ref{fig-gnky-irreducible} to the invariant amplitudes is
straightforward and explicit expressions are listed in~\cite{Sal03}. 

\begin{figure}[htbp]
\includegraphics[width=.7\textwidth]{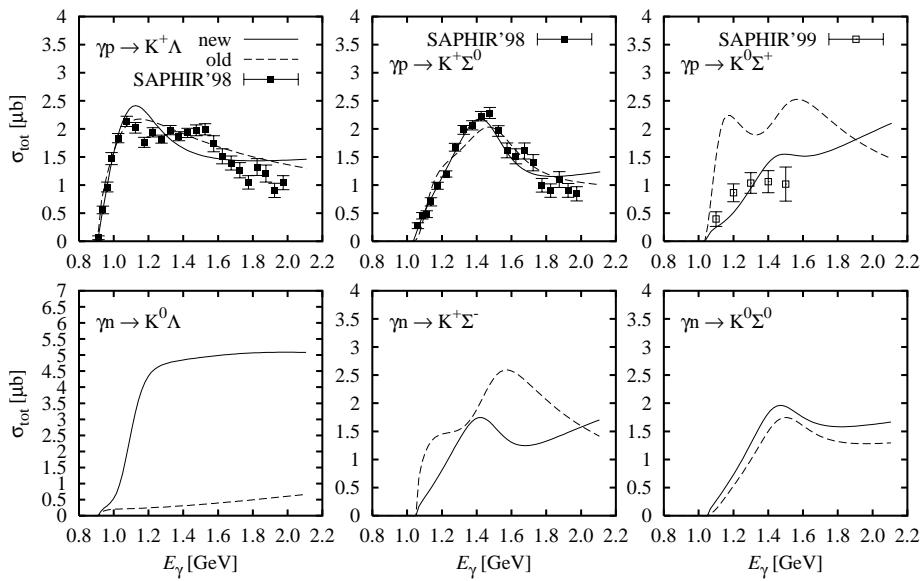}
\caption{Total cross sections of kaon photoproduction on
  the nucleon versus photon lab energy. Solid curves: 
  the model of~\cite{LeM01}; dashed curves: model of~\cite{BeM96};
  experimental data from the SAPHIR collaboration~\cite{Tra98,Goe99}.} 
\label{fig-resdisc-gnky-tc}
\end{figure}

The coupling constants and cut-off parameters were determined by a fit to
the experimental data. Fig.~\ref{fig-resdisc-gnky-tc} shows the
total cross section for the various channels as obtained from this
model together with experimental data~\cite{Tra98,Goe99} and with its 
older version~\cite{BeM96}. One readily notes
that the new model describes the data for $\gamma p\rightarrow K^+ \Sigma^0$
slightly better than the old one and considerably better for $\gamma {p}
\rightarrow {K}^{0}\Sigma^{+}$. However, the prediction for the
channel $\gamma {n}\rightarrow {K}^{0}\Lambda$ appears unrealistically
large. The authors of
\cite{LeM01} explain this feature by the lack of experimental data in
that channel making the parameter fitting uncontrollable. It may well
be that the older model may
give a more realistic description for this channel.
Nevertheless, we use the new model in the present work. 

%%%%%%%%%%%%%%%%%%%%%%%%%%%%%%%%%%%%%%%%%%%%%%%%%%%%%%%%%%%%%%%%%%%%%%%%%%%%%%%
\subsection{Pion photoproduction on the nucleon}\label{sub_gamma_pi}

\begin{figure}[htbp]
\includegraphics[width=.5\textwidth]{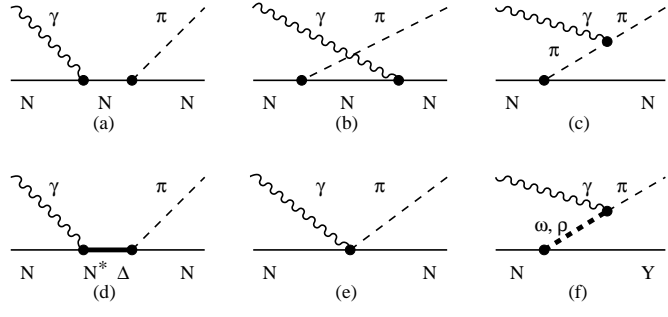}
\caption{Elementary diagrams for pion photoproduction on the
nucleon. Born terms: (a) - (c) nucleon, crossed nucleon and pion
poles and (e) Kroll-Rudermann contact term; (d): resonance term; (f):
vector meson exchange.} 
\label{fig-gnpn-irreducible}
\end{figure}

For pion photoproduction on the nucleon we use the MAID
model~\cite{DrH99}. The model contains Born terms (diagram (a)-(c),(e) of
Fig.~\ref{fig-gnpn-irreducible}), vector mesons $\rho$ and $\omega$ 
(diagram (f)), and a series of nucleon resonances $P_{33}(1232)$, 
$P_{11}(1440)$, $D_{13}(1520)$, $S_{11}(1535)$, $F_{15}(1680)$, 
and $D_{33}(1700)$ 
(diagram (d)). For the Born terms, this model uses both pseudoscalar and 
pseudovector coupling with a gradual transition from pure 
pseudovector coupling at threshold to pure pseudoscalar coupling
at high photon energies. This operator has been developed for photon 
energies up to 1.6~GeV which is above the threshold of kaon
photoproduction and, therefore, can be used for the evaluation of the
pion mediated reaction on the deuteron.  

\begin{figure}[htbp]
\includegraphics[width=.6\textwidth]{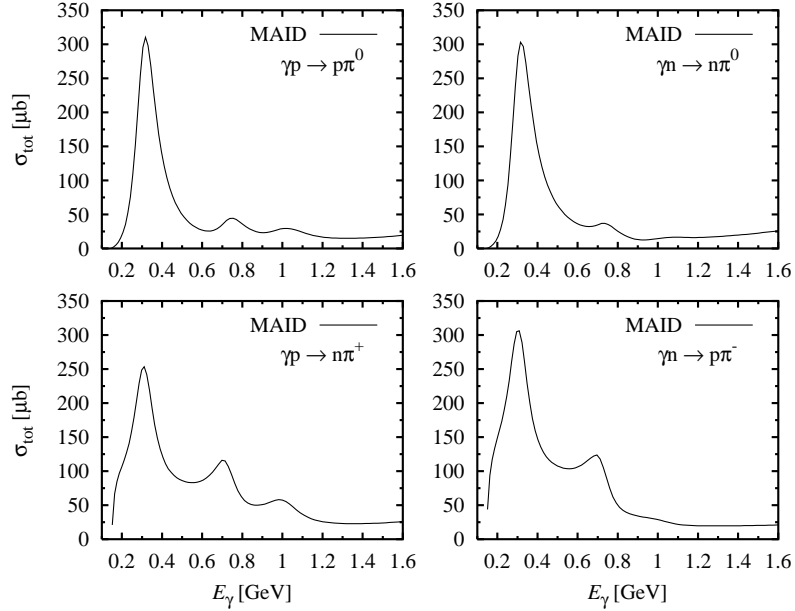}
\caption{Total cross sections of pion photoproduction on
  the nucleon versus photon lab energy from MAID~\cite{DrH99}.} 
\label{fig-resdisc-gnpn-tc}
\end{figure}

The resulting total cross sections are shown in
Fig.~\ref{fig-resdisc-gnpn-tc}. Comparing
Fig.~\ref{fig-resdisc-gnky-tc} with Fig.~\ref{fig-resdisc-gnpn-tc},
one clearly notes that the cross section for pion photoproduction on
the nucleon of about 25 $\mu$b around $E_\gamma=1.6$~MeV 
is still much stronger (about 10 times) than the cross section 
for kaon photoproduction at the same energies. This fact
indicates that the pion mediated photoproduction contribution may have
a sizeable influence on kaon photoproduction on the deuteron. 

The MAID-operator is given in the photon-pion c.m.\ system in terms of
the CGLN-amplitudes $F_{1}$ through $F_{4}$~\cite{ChG57} 
\begin{eqnarray}
{\cal M}^{\gamma \pi N}_{\mu_{N}^{\prime}\mu_{N}\lambda} &=&
\chi_{\mu_{N}^{\prime}}^{\dagger} 
(\vec \sigma \cdot \vec \epsilon_{\lambda}\, F_{1}
-i \vec \sigma \cdot \hat q\, 
\vec \sigma \cdot \hat k \times \vec \epsilon_{\lambda}\, F_{2}
+\vec \sigma \cdot \hat k\, 
\hat q \cdot \vec \epsilon_{\lambda}\, F_{3}
+\vec \sigma \cdot \hat q\, \hat q \cdot \vec
\epsilon_{\lambda}\, F_{4})
\chi_{\mu_{N}}\,,
\label{eq-gnpn-M-matrix-F}
\end{eqnarray}
where $\hat a$ means a unit vector in the direction of $\vec a$, $\vec
\sigma$ denotes the nucleon spin operator. The $F_i$ amplitudes are
functions of the invariant mass $W$ and the relative c.m.\ angle $\theta$ 
between photon and pion momentum. Since,
however, we need the amplitude in a general frame of reference for the
process on the deuteron, we have to establish a relation connecting
the $F_i$ amplitudes, defined in the c.m.\ system, to the invariant 
amplitudes $A_i^{\pi\gamma}$, defined in analogy to
Eq.~(\ref{eq-gnky-M-matrix-Dirac-spinor}), in any system. Comparing the
two representations, one finds as desired relation 
\begin{eqnarray}
\left(\begin{array}{c}
A^{\gamma \pi N}_1 \\ A^{\gamma \pi N}_2 \\ A^{\gamma \pi N}_3 \\ 
A^{\gamma \pi N}_4
\end{array}\right) &=& 4\pi 
\left(\begin{array}{cccc}
 W^+ &-W^- &-2m_{N} \frac{q \cdot k}{W^-} &
 2m_{N} \frac{q \cdot k}{W^+} \\
 0 & 0 & 1 &-1 \\
 1 & 1 & -W^+\frac{q \cdot k}{W^-} & -W^-\frac{q \cdot k}{W^+} \\
 1 & 1 & -\frac{q \cdot k}{W^-} & -\frac{q \cdot k}{W^+}
\end{array}\right)
\left(\begin{array}{c}
\frac{1}{\sqrt{E_{N}^{\prime\,+}E_{N}^+}}\frac{1}{W^-} F_{1} \\
\frac{1}{\sqrt{E_{N}^{\prime\,-}E_{N}^-}}\frac{1}{W^+} F_{2} \\
\frac{1}{qk}\sqrt{\frac{E_{N}^+}{E_{N}^{\prime\,+}}}
\frac{1}{W^+} F_{3} \\
\frac{1}{qk}\sqrt{\frac{E_{N}^-}{E_{N}^{\prime\,-}}}\frac{1}{W^-} F_{4}
\end{array}\right)\,,
\label{eq-gnpn-A-F-amplitude}
\end{eqnarray}
where we have introduced the notation $W^\pm=W\pm m_{N}$ and
analogously for $E_{N}^\pm$ and $E_{N}^{\prime\,\pm}$.
We would like to point out that the notation $q$ and $k$
in Eq.~(\ref{eq-gnpn-A-F-amplitude}) means the absolute value of
three-vectors whereas $q\cdot k$ refers to the 4-vector scalar
product. With these relations, we can evaluate 
${\cal M}^{\gamma \pi N}$ in any frame because
of the invariant character of the $A^{\gamma \pi N}_i$. 

%%%%%%%%%%%%%%%%%%%%%%%%%%%%%%%%%%%%%%%%%%%%%%%%%%%%%%%%%%%%%%%%%%%%%%%%%%%%%%%
\subsection{Hyperon-nucleon scattering}\label{subYN}

\begin{figure}[htbp]
\includegraphics[width=.5\textwidth]{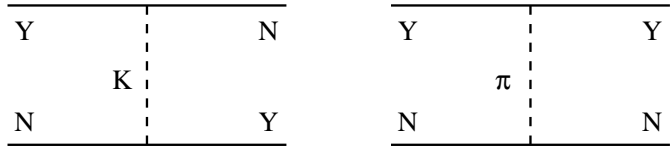}
\caption{\small Boson exchange diagrams of the hyperon-nucleon
potential $V_{YN}$. The left diagram describes a strangeness exchange and 
the right one a non-strangeness exchange.} 
\label{fig-ynyn-obep}
\end{figure}

For hyperon-nucleon scattering we use the Nijmegen interaction 
potential $V_{YN}$ from~\cite{MaR89,RiS99}. This interaction is described in
terms of one-boson exchanges. Since the hyperon is a baryon with
strangeness $S=-1$, the exchanges contain both strange and non-strange
mesons. The basic diagrams of this model are displayed in
Fig.~\ref{fig-ynyn-obep}. 
Three types of mesons are considered, pseudoscalar, scalar, and
vector mesons. The pseudoscalars are $\pi$, $\eta$, $\eta^{\prime}$, and K
with an $\eta-\eta^{\prime}$ mixing angle $\theta_P=-23.0^\circ$
from the Gell-Mann-Okubo mass formula. As vector mesons are included
$\rho$, $\phi$, K$^*$, and $\omega$ with an ideal $\phi-\omega$ 
mixing angle $\theta_V=37.56^\circ$ and as scalar mesons $\delta$,
$S^*$, $\kappa$, and $\varepsilon$ with a free
$S^*-\varepsilon$ mixing angle to be determined in a fit to the
$YN$ scattering data. Also included are contributions from Pomeron, $f$,
$f^{\prime}$, and $A_2$ exchanges. For details we refer
to~\cite{MaR89,RiS99}. 

%%%%%%%%%%%%%%%%%%%%%%%%%%%%%%%%%%%%%%%%%%%%%%%%%%%%%%%%%%%%%%%%%%%%%%%%%%%%%%%
\subsection{Kaon-nucleon scattering}

\begin{figure}[htbp]
\includegraphics[width=.7\textwidth]{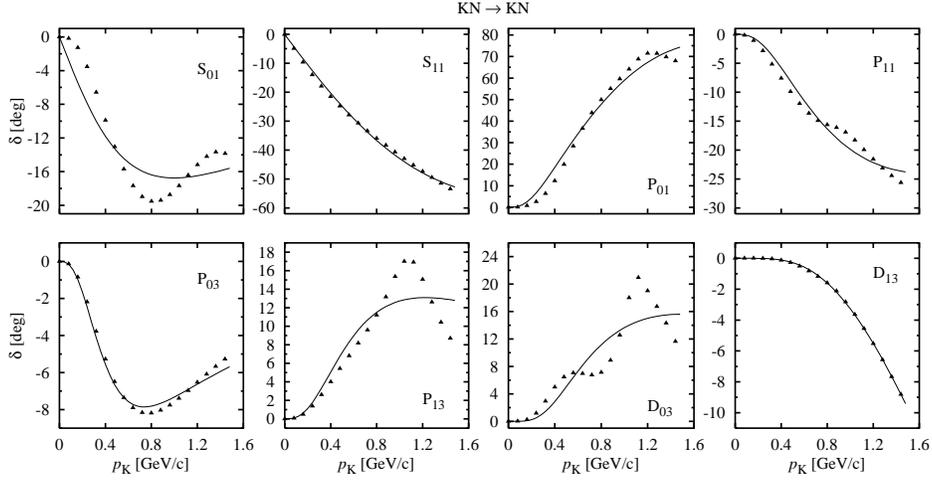}
\caption{Phase shifts of kaon-nucleon scattering versus
kaon lab momentum. Solid curves: results for a separable
potential of rank-1; triangles: phase shifts from the 
SAID-analysis~\cite{HyA92}.} 
\label{fig-resdisc-knkn-ps}
\end{figure}

In order to estimate the supposedly smaller influence of kaon-nucleon
rescattering, we take a simple rank-one separable interaction
potential for which the partial wave representation reads 
\begin{eqnarray}
V_{KN}^{\ell J}(p^{\prime},p) &=&
\lambda^{\ell J}_{KN}\, g^{\ell J}_{KN}(p^{\prime})\, 
g^{\ell J}_{KN}(p)\,,
\label{eq-knkn-V-separable-rank-1}
\end{eqnarray}
where $\lambda^{\ell J}_{KN}=\pm 1$ is a phase parameter and
$g^{\ell J}_{KN}(p)$ is a form factor. It is taken in the form 
\begin{eqnarray}
g^{\ell J}_{KN}(p) &=& \frac{B^{\ell J}_{KN}\, p^{\ell}}
{\left[p^{2}+(A^{\ell J}_{KN})^{2}\right]^{\frac{\ell+2}{2}}}\,,
\label{eq-knkn-form-factor}
\end{eqnarray}
where $B^{\ell J}_{KN}$ and $A^{\ell J}_{KN}$ are 
parameters which
characterize strength and range of the potential. 
\begin{table}[htbp]
\begin{center}
\caption{Parameters of the separable potential of rank-1 for
  kaon-nucleon scattering.} 
\label{tab-resdisc-knkn-pw}
\renewcommand{\arraystretch}{1.3}
\begin{ruledtabular}                                                 
\begin{tabular}{ccccc}
partial wave & I & $\lambda$ & $A^{\ell J} $ [MeV] & $B^{\ell J}$ [MeV]\\
\hline 
$S_{01}$ & 0 & + & 617.56 & 431.84 \\
$P_{01}$ & 0 & $-$ & 908.53 & 1815.2 \\
$P_{03}$ & 0 & + & 353.27 & 204.55 \\
$D_{03}$ & 0 & $-$ & 513.34 & 572.94 \\
\hline 
$S_{11}$ & 1 & + & 763.32 & 1049.9 \\
$P_{11}$ & 1 & + & 547.25 & 630.43 \\
$P_{13}$ & 1 & $-$ & 569.42 & 471.68 \\
$D_{13}$ & 1 & + & 1256.0 & 4979.8 \\
\end{tabular}
\end{ruledtabular}
\end{center}
\end{table}
With this
potential, the Lippmann-Schwinger equation for the partial wave
${\cal T}$-matrix reads
\begin{eqnarray}
{\cal T}^{\ell J}_{KN}(W;p^{\prime},p) &=&\lambda^{\ell J}_{KN}\,
g^{\ell J}_{KN}(p^{\prime})\,\left( g^{\ell J}_{KN}(p)
+ 2m_{KN}\int_{0}^{\infty} dp^{\prime\prime}\, (p^{\prime\prime})^{2}\,
\frac{g^{\ell J}_{KN}(p^{\prime\prime})\, 
T ^{\ell J}_{KN}(W,p^{\prime\prime},p)}
{q^{2}-(p^{\prime\prime})^{2}+i\varepsilon}\right),
\label{eq-knkn-T-matrix-separable-rank1}
\end{eqnarray}
which can be solved analytically yielding 
\begin{eqnarray}
T ^{\ell J}_{KN}(W,p^{\prime},p) &=&
\frac{\lambda^{\ell J}_{KN}\,g^{\ell J}_{KN}(p^{\prime})\, g^{\ell J}_{KN}(p)}
{1- 2\,m_{KN}\,\lambda^{\ell J}_{KN} \int_{0}^{\infty} 
dp^{\prime\prime}\, (p^{\prime\prime})^{2}\,
\frac{\left(g^{\ell J}_{KN}(p^{\prime\prime})\right)^{2}}
{q^{2}-(p^{\prime\prime})^{2}+i\varepsilon}}\,.
\label{eq-kn-T-matrix-separable-rank1-final}
\end{eqnarray}
For $\ell=0$ one finds explicitly
\begin{eqnarray}
{\cal T}^{0J}_{KN}(p^{\prime},p) &=&
\frac{\lambda^{0J}_{KN}\,(B^{0J}_{KN})^{2}}
{\left[(p^{\prime})^{2}+(A^{0J}_{KN})^{2}\right]
 \left[p^{2}+(A^{0J}_{KN})^{2}\right]}
\left[1+\frac{\pi m (B^{0J}_{KN})^{2}}
{2 A^{0J}_{KN} \left[A^{0J}_{KN}-iq\right]^{2}}\right]^{-1},
\label{eq-knkn-T-matrix-separable-rank1-0}
\end{eqnarray}
where the parameters of the potential are determined by fitting the
$(\ell=0)$-phase shifts to experimental data. The results for the
phase shifts are shown in~Fig.~\ref{fig-resdisc-knkn-ps} together with
the phase shifts of the SAID-analysis~\cite{HyA92} based on experimental 
data. The resulting parameters of this fit 
are listed in Table~\ref{tab-resdisc-knkn-pw}. 
\begin{figure}[htbp]
\includegraphics[width=.8\textwidth]{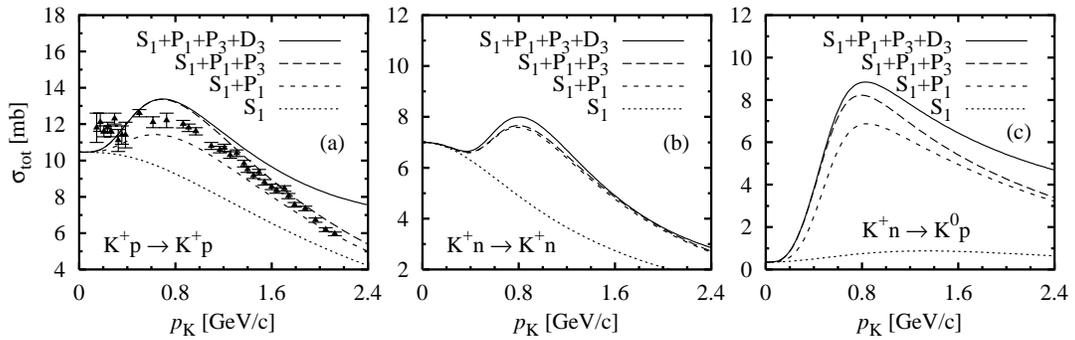}
\caption{Total elastic cross sections of various channels
  for kaon-nucleon scattering versus kaon lab momentum. Curves:
  predictions using the separable potential of rank-1; 
  experimental data from~\cite{Cam74,Abe75}. Notation S$_{1}$
  means S$_{01}$+S$_{11}$ and similarly for P$_{1}$, P$_{3}$ and
  D$_{3}$.} 
\label{fig-resdisc-knkn-tc}
\end{figure}

Fig.~\ref{fig-resdisc-knkn-ps} shows that the separable potential 
of rank-1 describes most of the phase shifts fairly well. 
The best fit is achieved for the
partial wave D$_{13}$ because of its simple form. Relatively good fits
are also obtained for S$_{11}$, P$_{01}$, P$_{11}$, and P$_{03}$. The
oscillatory form of D$_{03}$ and S$_{01}$ and the relatively sharp peak of
P$_{13}$ can not be fitted well by this simple separable potential. 
However, for the present study of the influence of $KN$ rescattering
this fit is good enough.

The total elastic cross sections of kaon-nucleon scattering with the
increasing contribution of partial waves are shown in
Fig.~\ref{fig-resdisc-knkn-tc}. The left panel (a) shows the cross
section for $K ^{+}p\rightarrow K ^{+}p$ having total isospin
$I=1$. The middle and right panels (b) and (c) show the cross sections
for channels having contributions from both total isospins $I=1$ and
$I=0$, i.e.\ $K ^{+} n\rightarrow K ^{+} n$ (panel (b)) and $K ^{+}
n\rightarrow K ^{0}p$ (panel (c)). For the reaction $K ^+p \rightarrow
K ^+p$ (panel (a)) a reasonable agreement with the data
from~\cite{Cam74,Abe75} is achieved by including S- and
P-waves only. D-waves show some effect above kaon momenta
$p_K\approx 1.6$~GeV/c.

%%%%%%%%%%%%%%%%%%%%%%%%%%%%%%%%%%%%%%%%%%%%%%%%%%%%%%%%%%%%%%%%%%%%%%%%%%%%%%%
\subsection{The $\pi N\rightarrow KY$ process}

The interaction $\pi N\rightarrow KY$ couples
the $\pi N$-channel with the $KY$-channel and thus one deals with a 
coupled two-channel problem. Therefore, the potential and the reaction 
matrix are described by $2\times 2$-matrices. Again we take for
simplicity a separable interaction potential, which reads in this case 
\beq
V_{KY,\pi N}(p',p)=\lambda g(p')_{KY,\pi N}\,g_{KY,\pi N}^\dagger(p)\,,
\eeq
where
\beq
g_{KY,\pi N}(p)=\left(\begin{array}{c}
g_{\pi N}(p)\\
g_{KY}(p)\\
\end{array}\right)
\eeq
with an analogous functional form for $g_{\pi N}(p)$ and $g_{KY}(p)$ as 
in~(\ref{eq-knkn-form-factor}). 
\begin{table}[htbp]
\caption{Parameters of the separable potential of rank-1 for
  the reaction $\pi N$ $\rightarrow$ KY with S-waves only.} 
\label{tab-resdisc-pnky-pw}
\renewcommand{\arraystretch}{1.3}
\begin{ruledtabular}                                                 
\begin{tabular}{cccccccc}
channel & partial wave & $\lambda$ & $A^{\ell J} $ [MeV] & 
$B^{\ell J}$ [MeV] \\
\hline 
$\pi N$ & $S_{11}$ & + & 1039.6& 559.03\\
$\pi N$ & $S_{31}$ & + & 140.97& 108.94\\
\hline 
$KY$ & $S_{11}$ & + & 179.39& 148.80\\
$KY$ & $S_{31}$ & + & 115.80& 224.41\\
\end{tabular}
\end{ruledtabular}
\end{table}
\begin{figure}[htbp]
\includegraphics[width=.7\textwidth]{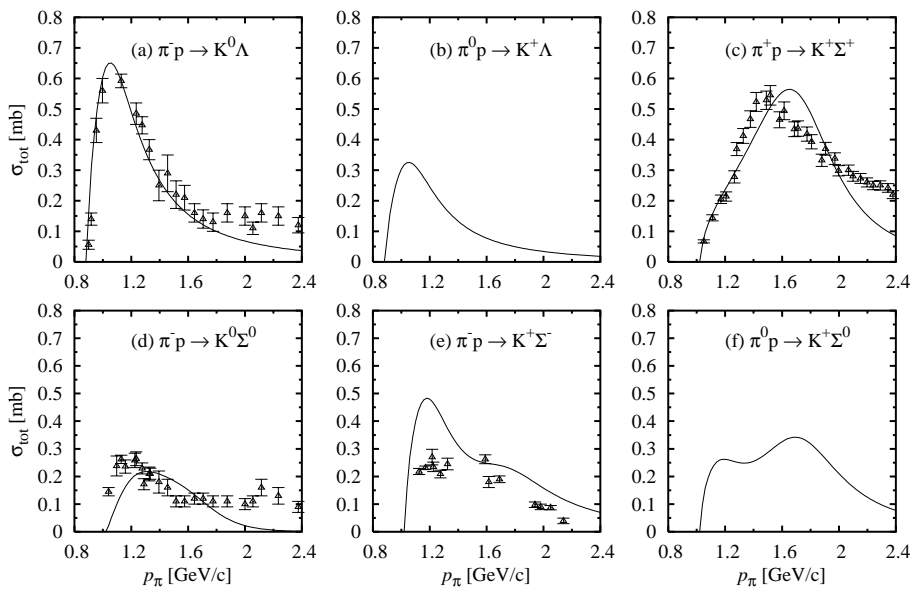}
\caption{Total cross sections of various channels for the
$\pi N$$\rightarrow$KY process versus pion lab-momentum. Solid
curves: calculations using a separable potential of rank-1 only in 
S-waves; triangles: experimental data taken from
  \cite{Kna75,Sax79,Bak78,Har79,Can83,Goo69,Dah67}.} 
\label{fig-resdisc-pnky-tc}
\end{figure}
The same steps like in kaon-nucleon scattering lead to the final form 
of the separable $T $-matrix 
\begin{eqnarray}
{\cal T}_{KY,\pi N}^{\ell J}(p^{\prime},p) &=&
\lambda\,g_{KY,\pi N}^{\ell J}(p^{\prime})\,(g^{\ell J}(p))_{KY,\pi N}^\dagger
\left[1-2\,\lambda \int_{0}^{\infty} 
dp^{\prime\prime}\, (p^{\prime\prime})^{2}\sum_{n\in\{\pi N,KY\}}
m_{n}\frac{\left(g^{\ell J}_{n}(p^{\prime\prime})\right)^{2}}
{q_{n}^{2}-(p^{\prime\prime})^{2}+i\varepsilon}\right]^{-1}\,.
\label{eq-pnky-T-matrix-separable-rank1-final}
\end{eqnarray}
For $\ell=0$ the $\pi N\rightarrow KY$ transition amplitude is given by 
\begin{eqnarray}
{\cal T}_{KY,\pi N}^{0J}(p^{\prime},p) &=&
\frac{B^{0J}_{\pi N}B^{0J}_{KY}}
{\left[(p^{\prime})^{2}+(A^{0J}_{\pi N})^{2}\right]
 \left[p^{2}+(A^{0J}_{KY})^{2}\right]} 
\left[1
+\pi \sum_{n\in\{\pi N,KY\}}
\frac{m_{n} (B^{0J}_{n})^{2}}
{2 A^{0J}_{n} \left[A^{0J}_{n}-iq_{n}\right]^{2}}
\right]^{-1}\,.
\label{eq-pnky-T-matrix-separable-rank1-0}
\end{eqnarray}
The parameters of the transition potential, listed in
Table~\ref{tab-resdisc-pnky-pw}, are determined by fitting the  
cross section $\pi N\rightarrow KY$ to experimental data. 

The total cross sections for the reaction $\pi N$ $\rightarrow$ KY are
shown in Fig.~\ref{fig-resdisc-pnky-tc}. Panels (a) and (b) are for the
$\Lambda$-channels and the others for the $\Sigma$-channels. The
experimental data for channel $K ^0\Lambda$ are taken from
\cite{Kna75,Sax79}, for $K ^+\Sigma^+$ from \cite{Can83}, for
$K ^0\Sigma^0$ from \cite{Bak78,Har79}, and for $K ^+\Sigma^-$ from
\cite{Goo69,Dah67}. The solid lines are obtained by fitting the cross
sections calculated by using a separable potential of rank-1 to the
available experimental data. In this fit we take into account only 
S-waves with total isospin $I=\frac{1}{2}$ and $I=\frac{3}{2}$, while
the addition of D-waves does not improve the convergence with
increasing kaon momentum. The resulting values of the potential
parameters are given in Table \ref{tab-resdisc-pnky-pw}. 

In Fig.~\ref{fig-resdisc-pnky-tc} it is seen that the separable
potential of rank-1 with S-waves only can fit experimental data
relatively well, especially in the channel $\pi^-$p $\rightarrow K ^0
\Lambda$. In this channel, the theoretical calculation underpredicts
experimental data at kaon lab-momenta above about 1.6 GeV/c. This
indicates that the higher partial waves become important in this energy
region. Also it can be seen in the panel (c) that the separable
potential of rank-1 with S-waves only cannot fit the data very well
above about 1.6 GeV/c. However, as a reasonable approximation we keep 
only the S-waves in the calculation. 

%%%%%%%%%%%%%%%%%%%%%%%%%%%%%%%%%%%%%%%%%%%%%%%%%%%%%%%%%%%%%%%%%%%%%%%%%%%%%%%
\section{Kaon photoproduction on the deuteron}
\label{sec3}

Now we will turn to kaon photoproduction on the deuteron 
\begin{eqnarray}
\gamma(p_{\gamma}) + {d}(p_{d})
&\rightarrow&
{K}(p_{K}) + {Y}(p_{Y}) + {N}(p_{N})\,,
\label{eq-gdkyn}
\end{eqnarray}
where $p_{\gamma}$, $p_{d}$, $p_{K}$, $p_{Y}$, and $p_{N}$ denote 
the 4-momenta of photon, deuteron, kaon, hyperon, and
nucleon, respectively. We begin with a brief review of the general 
formalism for cross section and target asymmetries. 

The general expression for the unpolarized cross section according 
to~\cite{BjD64} is given by
\begin{eqnarray}
d\sigma
&=&
\frac{\delta^{4}(p_{\gamma}+p_{d}
-p_{K}-p_{Y}-p_{N})m_{N}m_{Y}d^{3} p_{N}d^{3} p_{Y}d^{3} p_{K}}
{48(2 \pi)^{5} |\vec v_{\gamma} - \vec v_{d}|E_{\gamma}E_{d}E_{N}E_{Y}E_{K}}
 \sum_{\mu_{Y}\mu_{N}\mu_{d}\lambda} 
\left |{\cal M}^{K\gamma d}_{\mu_{Y}\mu_{N}\mu_{d}\lambda}
(\vec p_{Y},\vec p_{N},\vec p_{K},
\vec p_{d},\vec p_{\gamma}) \right|^{2}\,,
\label{eq-gdkyn-dsigma}
\end{eqnarray}
where $\mu_{Y}$, $\mu_{N}$, $\mu_{d}$, and $\lambda$ denote
the spin projections of hyperon, nucleon, deuteron and the photon
polarization, respectively. Covariant state normalization in the 
convention of~\cite{BjD64} is assumed. 

\begin{figure}[htbp]
\includegraphics[width=.65\textwidth]{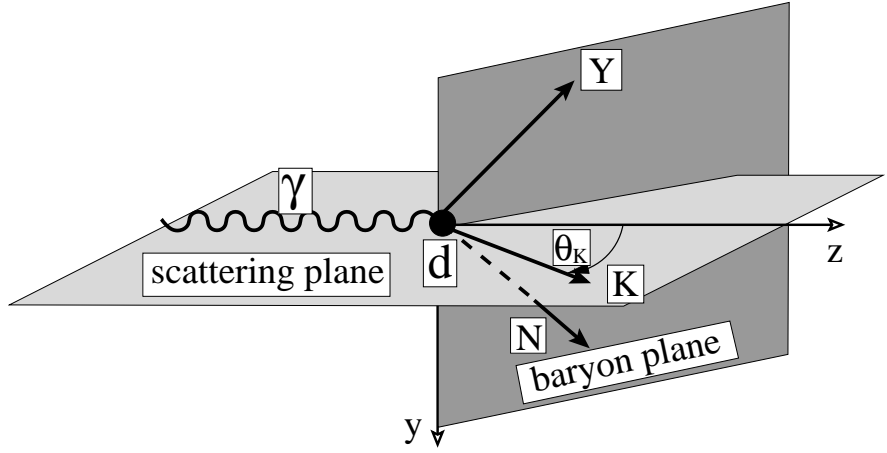}
\caption{\small The kinematics of kaon photoproduction on the deuteron
  in the lab frame.} 
\label{fig-gdkyn-kinematic}
\end{figure}

This expression is evaluated in the lab or deuteron rest frame. 
We have chosen a right-handed coordinate system where the
$z$-axis is defined by the photon momentum $\vec p_{\gamma}$ and the
$y$-axis by $\vec p_{\gamma} \times \vec p_{K}$. The kinematical 
situation is shown in Fig.~\ref{fig-gdkyn-kinematic}. The scattering
plane is defined by the momenta of photon $\vec p_\gamma$ and kaon 
$\vec p_K$ whereas the momenta of nucleon 
$\vec p_N$ and hyperon $\vec p_Y$ define the baryon plane.
In this frame the kinematical variables of the initial state are 
\begin{eqnarray}
p_{d} &=& (M_d,\,\vec 0) \quad \mbox{and}\quad
p_{\gamma} = (p_{\gamma},\,p_{\gamma} \hat z)\,.
\label{eq-gdkyn-deuteron-photon-momentum}
\end{eqnarray}
The threshold lab energy is given by
\beq
p_{\gamma}^{thr}=\frac{(m_{YN}+m_K)^2-m_d^2}{2m_d}\,,
\eeq
where $m_{YN}=m_Y+m_N$. 
For the final state, we choose as independent variables the kaon 
three-momentum $\vec p_{K}=(p_{K},\theta_{K},\phi_{K})$, where we can
choose $\phi_{K}=0$ since we do not consider polarized
photons. Furthermore, the spherical angles 
$\Omega^{\,\ast}_{YN}=\hat
p^{\,\ast}_{YN}=(\theta^{\,\ast}_{YN},\phi^{\,\ast}_{YN})$ 
of the relative $YN$-momentum $\vec p^{\,\ast}_{YN}=(\vec
p^{\,\ast}_{Y}-\vec p^{\,\ast}_{N})/2$ in the $YN$-c.m.\ system as
indicated by the asterisk. The relation to the corresponding 
lab frame quantities is obtained by an appropriate Lorentz boost with
$\vec \beta=\vec P_{YN}/E_{YN}$, where
\beq
\vec P_{YN}=\vec p_Y+\vec p_N=\vec p_\gamma -\vec p_K\,,\quad
\mbox{and}\quad E_{YN}=E_Y+E_K=E_\gamma +m_d-E_K\,.
\eeq

The orientation of the baryon plane is characterized
by $\Omega_{YN}$, the spherical angles of the relative $YN$-momentum in
the lab system. For given photon energy and kaon emission angle $\theta_K$,
the kaon momentum $p_K$ is bounded by $p_K^{min}\leq p_K\leq p_K^{max}$. In 
order to determine the boundaries we consider first the kaon momentum
for a given invariant mass $W_{YN}=\sqrt{E_{YN}^2-\vec P_{YN}^{\,2}}$ 
of the $YN$-subsystem yielding two solutions
\beq
p_K=\frac{1}{2b}\,\Big(a\,p_\gamma \cos \theta_K 
\pm E_{\gamma d}\sqrt{a^2-4b^2m_K^2}\Big)\,,\label{kaon_momentum}
\eeq
where
\beqa
a&=&W_{\gamma d}^2+m_K^2-W_{YN}^2\,,\\
b&=&W_{\gamma d}^2+p_\gamma^2\sin^2\theta_K\,,\\
W_{\gamma d}^2&=&E_{\gamma d}^2-p_\gamma^2=m_d(m_d+2\,p_\gamma)\,,\\
E_{\gamma d}&=&m_d+p_\gamma\,.
\eeqa
The upper and lower limits of $p_K$ are determined by the minimal value of
$W_{YN}^2$ which is $W_{YN}^2= m_{YN}^2$, resulting in 
\beqa
p_K^{max}&=& \frac{1}{2b}\,\Big(a_0\,p_\gamma \cos \theta_K 
+ E_{\gamma d}\sqrt{a_0^2-4b^2m_K^2}\Big)\,,\\
p_K^{min}&=& \max\{0,\frac{1}{2b}\,\Big(a_0\,p_\gamma \cos \theta_K 
- E_{\gamma d}\sqrt{a_0^2-4b^2m_K^2}\Big)\}\,,
\eeqa
where $a_0=W_{\gamma d}^2+m_K^2-m_{YN}^2$. The lower limit 
$p_K=(a_0\,p_\gamma \cos \theta_K- E_{\gamma
d}\sqrt{a_0^2-4b^2m_K^2})/2b>0$ applies if $0\le \theta_K\le \pi/2$
and, since $a_0>0$, if
\beq
a_0\,p_\gamma \cos \theta_K > E_{\gamma d}\sqrt{a_0^2-4b^2m_K^2}\,,
\eeq
which happens for 
\beq
p_\gamma> (m_{YN}^2-(m_d-m_K)^2)/(2(m_d-m_K))\,.
\eeq

Integrating over the kaon momentum $p_k$ and over
$\Omega^{\,\ast}_{YN}$, one obtains the semi-inclusive differential
cross section of kaon photoproduction on the deuteron, where only the
final kaon is detected without analyzing its energy, 
\begin{eqnarray}
\frac{d\sigma}{d\Omega_{K}} 
&=& \int_{p_K^{min}}^{p_K^{max}} dp_{K} \int
d\Omega^{\,\ast}_{YN}\, \kappa\,  
 \sum_{\mu_{Y}\mu_{N}\mu_{d}\lambda} 
\left \vert {\cal M}^{K\gamma d}_{\mu_{Y}\mu_{N}\mu_{d}\lambda}
(\vec p_{YN},\,\vec p_{K},\,\vec p_{\gamma}) \right \vert^{2}
\label{eq-gdkyn-dsigma/domegak}
\end{eqnarray}
with a kinematic factor
\begin{eqnarray}
\kappa &=& \frac{m_{Y} m_{N} p_{K}^{2} p^{\,\ast}_{YN}}
{24(2\pi)^{2} p_{\gamma}E_{K} W_{YN}}\,.
\label{eq-gdkyn-kinematical-factor}
\end{eqnarray}

With respect to polarization observables, we consider only the tensor
target asymmetries $T_{2M}$ which we define in analogy to deuteron
photodisintegration~\cite{Are88} writing the differential cross section 
for a tensor polarized deuteron target in the form
\beqa
\frac{d\sigma(P_2^d)}{d\Omega_{K}} &=&\frac{d\sigma}{d\Omega_{K}} 
\Big(1+P_2^d\sum_{M=0}^2 T_{2M}(\theta_K)\cos[M(\phi_K-\phi_d)]\,
d^2_{M0}(\theta_d)\Big)\,,
\eeqa
where $d^2_{M0}(\theta_d)$ denotes a small rotation matrix~\cite{Edm57} 
and $P_2^d$ the degree of tensor polarization with respect 
to an orientation axis with spherical angle $(\theta_d, \phi_d)$. 
The tensor polarization parameter is defined by $P_2^d=(1-3p_0)/\sqrt{2}$,
where $p_0$ denotes the probability to find a deuteron spin projection 
$m_d=0$ on the orientation axis. Then one has
\begin{eqnarray}
T_{2M} \frac{d\sigma}{d\Omega_{K}} &=&
(2-\delta_{M0})\, {\mathcal Re}\, V_{2M}\,,\quad M=0, 1, 2\,,
\label{eq-gdkyn-T2m}
\end{eqnarray}
where
\begin{eqnarray}
V_{2M} &=& 
\sqrt{15} 
\sum_{\mu_{Y}\mu_{N}\lambda} 
\sum_{\mu^{\prime}_{d}\mu_{d}}
(-1)^{1-\mu^{\prime}_{d}}
\left(\begin{array}{ccc}
    1       &      1                &  2 \\
\mu_{d} & -\mu^{\prime}_{d} & -M 
\end{array}\right) 
\int_{p_K^{min}}^{p_K^{max}} dp_{K} \int d\Omega^{\,\ast}_{YN}\, \kappa 
({\cal M}^{K\gamma d}_{\mu_{Y}\mu_{N}\mu_{d}\lambda})^{*}
{\cal M}^{K\gamma d}_{\mu_{Y}\mu_{N}\mu^{\prime}_{d}\lambda}\,.
\label{eq-gdkyn-V2m}
\end{eqnarray}
We use the convention of Edmonds~\cite{Edm57} for the $3j$-symbols.

%%%%%%%%%%%%%%%%%%%%%%%%%%%%%%%%%%%%%%%%%%%%%%%%%%%%%%%%%%%%%%%%%%%%%%%%%%%%%%%
\subsection{The photoproduction amplitude}

\begin{figure}[htbp]
\includegraphics[width=.6\textwidth]{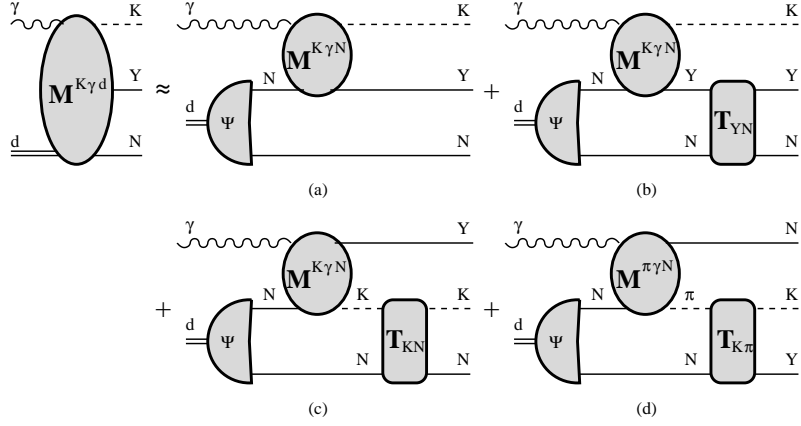}
\caption{\small Kaon photoproduction on the deuteron with 
rescattering contributions in the two-body subsystems and the pion
mediated process. Diagram (a): impulse approximation (IA);
(b) and (c): $YN$  and $KN$ rescattering, respectively; (d): 
$\pi N\rightarrow KY$ process.} 
\label{fig-gdkyn-expand-approx}
\end{figure}

All observables are determined by the photoproduction amplitude
${\cal M}^{K\gamma d}_{\mu_{Y}\mu_{N}\mu_{d}\lambda}$ which is the matrix 
element of a corresponding photoproduction operator 
$\widehat{\cal M}^{K\gamma d}$, i.e.
\beq
{\cal M}^{K\gamma d}_{\mu_{Y}\mu_{N}\mu_{d}\lambda}
(\vec p_{YN},\,\vec p_{K},\,\vec p_{\gamma})=
\langle \vec p_{YN}\,\vec p_{K}\,\mu_{Y}\,\mu_{N}|
\widehat{\cal M}^{K\gamma d}|\vec p_{\gamma}\,\mu_{d}\,\lambda\rangle\,. 
\eeq
In principle, the full treatment of all interaction effects requires 
a three-body treatment. In the present
work, however, we will restrict ourselves to the inclusion of complete
rescattering in the various two-body subsystems of the final
state. This is legitimate as a first step in order to see how
important interaction effects are at all. 
In this approximation, the production operator 
$\widehat{\cal M}^{K\gamma d}$ can be
written as  
\begin{eqnarray}
\widehat{\cal M}^{K\gamma d} &=& \widehat{\cal M}^{K\gamma d}_{IA} 
+ \widehat{\cal M}^{K\gamma d}_{YN} + \widehat{\cal M}^{K\gamma
d}_{KN} + \widehat{\cal M}^{K\gamma d}_{K\pi}\,, 
\label{eq-gdkyn-M-matrix}
\end{eqnarray}
where $\widehat{\cal M}^{K\gamma d}_{IA}$, 
$\widehat{\cal M}^{K\gamma d}_{YN}$, $\widehat{\cal M}^{K\gamma d}_{KN}$, 
and $\widehat{\cal M}^{K\gamma d}_{K\pi}$ denote the operators 
for the impulse
approximation, hyperon-nucleon rescattering, kaon-nucleon
rescattering, and the pion mediated process,
respectively. The graphical representation of this approximation is
displayed in Fig.~\ref{fig-gdkyn-expand-approx}. 

We will now describe the evaluation of each contribution in some detail. 

%%%%%%%%%%%%%%%%%%%%%%%%%%%%%%%%%%%%%%%%%%%%%%%%%%%%%%%%%%%%%%%%%%%%%%%%%%%%%%%

\subsubsection{The impulse approximation}

In the impulse approximation the incoming photon interacts with one nucleon 
only producing a kaon while the other nucleon remains untouched, i.e.\ 
acts merely as a spectator. Furthermore, any subsequent interaction is 
neglected, thus the final state is a pure plane wave. Then the
transition amplitude ${\cal M}^{K\gamma d}_{IA}$ for the diagram (a) of
Fig.~\ref{fig-gdkyn-expand-approx} is determined by the matrix element
of the elementary kaon photoproduction operator $\widehat{\cal
M}^{K\gamma N}$ taken between the deuteron bound state and the
three-particle plane wave final state 
\begin{eqnarray}
{\cal M}^{K\gamma d}_{IA,\,\mu_{Y}\mu_{N}\mu_{d}\lambda} &=&
\langle \vec p_{YN}\,\vec p_{K}\,\mu_{Y}\,\mu_{N}|\widehat
{\cal M}^{K\gamma N}|\vec p_{\gamma}\,\mu_{d}\,\lambda\rangle\,,
\label{eq-gdkyn-ia}
\end{eqnarray}
where the $\mu$'s denote the spin projection of the corresponding
particles and $\lambda$ the photon polarization. We use for the
$npd$-vertex 
\begin{eqnarray}
\langle \vec p_{N}\, \vec p_{N}^{\,\prime}\,1\,\mu_{S}  |
\vec p_{d}\, \mu_{d}\rangle &=&
(2\pi)^{3}\delta^{3}(\vec p_{d}-\vec p_{N}-\vec p_{N}^{\,\prime})
\frac{\sqrt{E_{N}E_{N}^{\prime}}}{m_{N}}
\Psi_{\mu_{S}\mu_{d}}(\vec p\,)\,,
\label{eq-gdkyn-two-nucleon-system}
\end{eqnarray}
where $\vec p=(\vec p_N-\vec p_{N}^{\,\prime})/2$ denotes the relative 
momentum of neutron and proton in the deuteron. The deuteron wave 
function has the form
\begin{eqnarray}
\Psi_{\mu_{S}\mu_{d}}(\vec p\,) &=&
\sqrt{2E_{d}(2\pi)^{3}}
\sum_{\ell=0,2} i^{\ell} C^{\ell 11}_{\mu_{\ell}\mu_{S}\mu_{d}} 
u_{\ell}(p) Y_{\ell\mu_{\ell}}(\hat p)\,,
\label{eq-gdkyn-deuteron-wave-function}
\end{eqnarray}
where $C$ denotes a Clebsch-Gordan coefficient. With $\vec p_{d}=0$ 
we obtain finally
\begin{eqnarray}
{\cal M}^{K\gamma d}_{IA,\,\mu_{Y}\mu_{N}\mu_{d}\lambda}
(\vec p_{YN},\,\vec p_{K},\,\vec p_{\gamma}) &=&
\sum_{\mu_{N^{\prime}}}
{\cal M}_{\mu_{Y}\mu_{N}\lambda}^{K\gamma N}
(\vec p_{Y},\vec p_{K},-\vec p_{N},\vec p_{\gamma}) 
C^{\frac{1}{2} \frac{1}{2} 1}
_{\mu_{N}\mu_{N^{\prime}} \mu_{N}+\mu_{N^{\prime}}}
\Psi_{\mu_{N}+\mu_{N^{\prime}}\mu_{d}}(-\vec p_{N})\,,
\label{eq-gdkyn-ia-final}
\end{eqnarray}
where $\vec p_Y=(\vec p_\gamma -\vec p_K)/2 +\vec p_{YN}$ and 
$\vec p_N=(\vec p_\gamma -\vec p_K)/2 -\vec p_{YN}$. This expression 
is straightforward to evaluate.

%%%%%%%%%%%%%%%%%%%%%%%%%%%%%%%%%%%%%%%%%%%%%%%%%%%%%%%%%%%%%%%%%%%%%%%%%%%%%%%

\subsubsection{$YN$ rescattering}

For the calculation of the $YN$ rescattering contribution, K.\ Miyagawa
made available to us the routine used in~\cite{YaM99}. This routine is
based on the combined evaluation of the diagrams (a) and (b) of
Fig.~\ref{fig-gdkyn-expand-approx}, i.e.\ the impulse 
approximation together with the subsequent hyperon-nucleon
rescattering contribution to the photoproduction operator, which is
written as 
\begin{eqnarray}
\widehat{\cal M}^{K\gamma d}_{IA+YN} &=& \widehat{\cal M}^{K\gamma d}_{IA} 
+ \widehat{\cal M}^{K\gamma d}_{YN} \\
&=& \widehat{\cal M}^{K\gamma d}_{IA} + 
\widehat{\cal T}_{YN}\, \widehat G_{YN} \widehat{\cal M}^{K\gamma d}_{IA}\,
\label{eq-gdkyn-yn-M-IA+YN}
\end{eqnarray}
with the $YN$-scattering operator $\widehat{\cal T}_{YN}$ and 
$\widehat G_{YN}$ as the free hyperon-nucleon off-shell 
propagator in the presence of a non-interacting kaon.
The scattering operator $\widehat{\cal T}_{YN}$ obeys the
Lippmann-Schwinger equation 
\begin{eqnarray}
\widehat{\cal T}_{YN} &=& \widehat{V}_{YN} + 
\widehat{V}_{YN} \,\widehat G_{YN}\, \widehat{\cal T}_{YN}\,, 
\label{eq-gdkyn-yn-R-YN-LS}
\end{eqnarray}
where $\widehat{V}_{YN}$ denotes the hyperon-nucleon potential
operator introduced in Sect.~\ref{subYN}. Inserting
Eq.~(\ref{eq-gdkyn-yn-R-YN-LS}) in Eq.~(\ref{eq-gdkyn-yn-M-IA+YN}), we
get 
\begin{eqnarray}
\widehat{\cal M}^{K\gamma d}_{IA+YN} &=& \widehat{\cal M}^{K\gamma d}_{IA} 
+ \widehat{V}_{YN}\, \widehat G_{YN} \widehat{\cal M}^{K\gamma d}_{IA+YN}\,, 
\label{eq-gdkyn-yn-M-IA+YN-LS}
\end{eqnarray}
which can be solved by inversion
\begin{eqnarray}
\widehat{\cal M}^{K\gamma d}_{IA+YN} &=& 
\left(\widehat 1 - \widehat V_{YN} \,\widehat G_{YN} \right)^{-1}
\widehat{\cal M}^{K\gamma d}_{IA}\,. 
\label{eq-gdkyn-yn-M-IA+YN-LS-solved}
\end{eqnarray}
After solving the last equation in the partial wave decomposition with
respect to the hyperon-nucleon subsystem, one
obtains the $YN$ rescattering amplitude by subtraction of the
impulse approximation
\begin{eqnarray}
{\cal M}^{K\gamma d}_{YN,\,\mu_{Y}\mu_{N}\mu_{d}\lambda}
(\vec p_{YN},\, \vec p_{K}, \,\vec p_{\gamma}) &=& 
\sum_{\ell SJ \mu_J} 
C^{\frac{1}{2} \frac{1}{2} S}
_{\mu_{Y}\mu_{N} \mu_{S}}\,
C^{\ell SJ}_{\mu_{\ell}\mu_{S}\mu_{J}}\,
Y_{\ell \mu_{\ell}}(\hat p_{YN})\, \nonumber \\&& \times
\left({\cal M}^{K\gamma d}_{IA+YN,\,\ell SJ,\mu_{d}\lambda}
(\vec p_{YN},\, \vec p_{K}, \, \vec p_{\gamma})
-{\cal M}^{K\gamma d}_{IA,\,\ell SJ,\mu_{d}\lambda}
(\vec p_{YN},\, \vec p_{K}, \,\vec p_{\gamma})
\right)\,.
\label{eq-gdkyn-yn-M-YN}
\end{eqnarray}

%%%%%%%%%%%%%%%%%%%%%%%%%%%%%%%%%%%%%%%%%%%%%%%%%%%%%%%%%%%%%%%%%%%%%%%%%%%%%%%
\subsubsection{$KN$ rescattering}

We evaluate the $KN$ rescattering contribution (diagram (c) of
Fig.~\ref{fig-gdkyn-expand-approx}) directly in contrast to $YN$
rescattering. The corresponding amplitude is given by 
\begin{eqnarray}
{\cal M}^{K\gamma d}_{KN,\,\mu_{Y}\, \mu_{N} \,\mu_{d}\, \lambda}
(\vec p_{YN},\,\vec p_{K}, \,\vec p_{\gamma}) &=& 
\langle \vec p_{YN}\,\vec p_{K}\,\mu_{Y}\,\mu_{N}|
\widehat{\cal T }_{KN} \widehat G_{KN} \widehat{\cal M}^{K\gamma N}
|\vec p_{\gamma} \,\mu_{d}\,\lambda \rangle\,,
\label{eq-gdkyn-kn}
\end{eqnarray}
where $\widehat{\cal T }_{KN}$ is the
kaon-nucleon scattering operator, $\widehat G_{KN}$ is the free kaon-nucleon
propagator in the presence of a non-interacting
hyperon. Straightforward evaluation yields 
\begin{eqnarray}
{\cal M}^{K\gamma d}_{KN,\,\mu_{Y}\, \mu_{N} \,\mu_{d}\, \lambda} 
(\vec p_{YN},\, \vec p_{K}, \,\vec p_{\gamma})&=&
\sum_{\mu_{N}^{\prime}} 
\int d^{3} p_{KN}^{\prime}
\sqrt{\frac{E_{N} E_{K}}
{E_{N}^{\prime} E_{K}^{\prime}}}
{\cal T}_{KN,\,\mu_{N}\,\mu_{N}^{\prime}}
(\vec p_{KN}, \,\vec p_{KN}^{\,\prime})
G_{KN}(E_{KN}-E_{KN}^{\prime}) \nonumber \\&& \times\,
{\cal M}^{K\gamma d}_{IA,\,\mu_{Y}\, \mu_{N}^{\prime} \,\mu_{d}\, \lambda}
(\vec p_{YN'},\,\vec p_{KN}^{\,\prime},\, \vec p_{\gamma}) \,,
\label{eq-gdkyn-kn-middle}
\end{eqnarray}
where $\vec p_{KN}$ and $\vec p_{KN}^{\,\prime}$ denote the relative
momenta and $E_{KN}$ and $E_{KN}^{\prime}$ the total energies of 
the final and intermediate kaon-nucleon states, respectively,
and the free propagator is given by
\begin{eqnarray}
G_{KN}(z) &=& \frac{1}{z+i\varepsilon} 
= \frac{\mathcal P}{z} - i\pi\delta(z)\,.
\label{eq-gdkyn-kn-propagator-KN}
\end{eqnarray}
Inserting this expression into Eq.~(\ref{eq-gdkyn-kn-middle}), one finds
as the final expression for the amplitude of the kaon-nucleon
rescattering contribution
\begin{eqnarray}
{\cal M}^{K\gamma d}_{KN}(\vec p_{YN},\, \vec p_{K}, \,\vec p_{\gamma})
&=& 2m_{KN} 
\sum_{\mu_{N}^{\prime}} {\mathcal P}
\int d^{3} p_{KN}^{\prime}
\sqrt{\frac{E_{N} E_{K}}{E_{N}^{\prime} E_{K}^{\prime}}}\,
\frac{{\cal T}_{KN,\,\mu_{N}\,\mu_{N}^{\prime}}
(\vec p_{KN}, \,\vec p_{KN}^{\,\prime})
{\cal M}^{K\gamma d}_{IA,\,\mu_{Y}\, \mu_{N}^{\prime} \,\mu_{d}\, \lambda}
(\vec p_{YN'},\,\vec p_{KN}^{\,\prime},\,\vec p_{\gamma})}
{q_{KN}^{2}-(p_{KN}^{\prime})^{2}} 
\nonumber \\&&\hspace*{-1.5cm}
-i\pi\, m_{KN} \,q_{KN} 
\sum_{\mu_{N}^{\prime}} 
\int d\Omega_{KN}^{\prime}
\sqrt{\frac{E_{N} E_{K}}{E_{N}^{\prime} E_{K}^{\prime}}}\,
{\cal T}_{KN,\,\mu_{N}\, \mu_{N}^{\prime}}
(\vec p_{KN}, \vec q_{KN}^{\,\prime})
{\cal M}^{K\gamma d}_{IA,\,\mu_{Y}\, \mu_{N}^{\prime} \,\mu_{d}\, \lambda}
(\vec p_{YN'},\,\vec q_{KN}^{\,\prime},\,\vec p_{\gamma})\,.
\label{eq-gdkyn-kn-final}
\end{eqnarray}
In this expression, one has 
$\vec p_{YN'}=(\vec p_{Y}-\vec p_{N}^{\,\prime})/2$ with 
$\vec p_{N}^{\,\prime}=(\vec p_{\gamma}-\vec p_Y\,)/2-\vec p_{KN}^{\,\prime}$,
and $\vec q_{KN}^{\,\prime}
=\{q_{KN},\Omega_{KN}^{\prime}\}$ with $q_{KN}$ given by 
\begin{eqnarray}
q_{KN} &=& \sqrt{2m_{KN}\left(E_{KN}
-\frac{P_{KN}^{2}}{2M_{KN}}-M_{KN} \right)}\,,
\label{eq-gdkyn-kn-q}
\end{eqnarray}
where $m_{KN}$ and $M_{KN}$ denote the reduced and total masses of 
the kaon-nucleon system, respectively.

%%%%%%%%%%%%%%%%%%%%%%%%%%%%%%%%%%%%%%%%%%%%%%%%%%%%%%%%%%%%%%%%%%%%%%%%%%%%%%%
\subsubsection{$\gamma d \rightarrow \pi NN \rightarrow KYN$ process}

The contribution of the diagram (d) of
Fig.~\ref{fig-gdkyn-expand-approx} has formally the same structure
than the foregoing kaon-nucleon rescattering amplitude. We just have
to replace in the final result of Eq.~(\ref{eq-gdkyn-kn-final}) the
kaon photoproduction amplitude by the pion photoproduction amplitude and the
$KN$ scattering amplitude by the $\pi N \rightarrow KY$ transition
amplitude, i.e.
\beqa
{\cal M}^{K\gamma d}_{IA,\,\mu_{Y}\,\mu_{N}^{\prime} \,\mu_{d}\,
\lambda}(\vec p_{YN'},\,\vec q_{KN}^{\,\prime},\,\vec p_{\gamma})
&\rightarrow&{\cal M}^{\pi\gamma d}_{IA,\,\mu_{N}\,
\mu_{N}^{\prime} \,\mu_{d}\, \lambda}
(\vec p_{NN'},\,\vec q_{\pi N}^{\,\prime},\,\vec p_{\gamma})\,,\\
{\cal T}_{KN,\,\mu_{N}\,\mu_{N}^{\prime}}
(\vec p_{KN}, \,\vec p_{KN}^{\,\prime})&\rightarrow&
{\cal T}_{K\pi,\,\mu_{Y}\,\mu_{N}^{\prime}}
(\vec p_{KN}, \,\vec p_{\pi N}^{\,\prime})\,,
\eeqa 
where ${\cal T}_{K\pi}$ and ${\cal M}^{\pi\gamma d}_{IA}$ denote the
transition matrices for the pion mediated kaon process and the impulse
approximation of pion photoproduction on the deuteron,
respectively. The final result is 
\begin{eqnarray}
{\cal M}^{K\gamma d}_{K\pi}(\vec p_{YN},\, \vec p_{K}, \,\vec p_{\gamma})
&=& 2m_{\pi N} \sum_{\mu_{N}^{\prime}} {\mathcal P}
\int d^{3} p_{\pi N}^{\prime}
\sqrt{\frac{E_{K} E_{Y}}{E_{\pi} E_{N}^{\prime}}}\,
\frac{{\cal T}_{K\pi,\,\mu_{Y}\,\mu_{N}^{\prime}}
(\vec p_{KY},\, \vec p_{\pi N}^{\,\prime})
{\cal M}^{\pi\gamma d}_{IA,\,\mu_{N}\, \mu_{N}^{\prime}\, \mu_{d}\, \lambda}
(\vec p_{NN'},\,\vec p_{\pi N}^{\,\prime},\,\vec p_{\gamma})} 
{q_{\pi N}^{2}-(p_{\pi N}^{\prime})^{2}} 
\nonumber \\&&\hspace*{-1.cm}
-i\pi m_{\pi N} q_{\pi N} 
\sum_{\mu_{N}^{\prime}} 
\int d\Omega_{\pi N}^{\prime}
\sqrt{\frac{E_{K} E_{Y}}{E_{\pi} E_{N}^{\prime}}}\,
{\cal T}_{K\pi,\,\mu_{Y}\,\mu_{N}^{\prime}}
(\vec p_{KY},\, \vec q_{\pi N}^{\,\prime})
{\cal M}^{\pi\gamma d}_{IA,\,\mu_{N}\, \mu_{N}^{\prime}\, \mu_{d}\, \lambda}
(\vec p_{NN'}, \,\vec q_{\pi N}^{\,\prime},\,
\vec p_{\gamma})\,,
\label{eq-gdkyn-pk-final}
\end{eqnarray}
where $\vec p_{NN'}=(\vec p_{N}-\vec
p_{N}^{\,\prime})/2$ with $\vec p_{N}^{\,\prime}=(\vec p_{\gamma}-\vec
p_N\,)/2-\vec p_{\pi N}^{\,\prime}$, and $\vec q_{\pi N}^{\,\prime} =
\{q_{\pi N},d\Omega_{\pi N}^{\prime}\}$ with $q_{\pi N}$  given by 
\begin{eqnarray}
q_{\pi N} &=& \sqrt{2m_{\pi N}\left(E_{KY}
-\frac{P_{\pi N}^{2}}
{2M_{\pi N}}-M_{\pi N} \right)}\,.
\label{eq-gdkyn-pk-q}
\end{eqnarray}
Again $m_{\pi N}$ and $M_{\pi N}$ denote the reduced and total masses of 
the pion-nucleon system, respectively.

%%%%%%%%%%%%%%%%%%%%%%%%%%%%%%%%%%%%%%%%%%%%%%%%%%%%%%%%%%%%%%%%%%%%%%%%%%%%%%%
\section{Results and discussion}
\label{sec4}

The three contributions from hyperon-nucleon and kaon-nucleon
rescattering and from the pion mediated process have been evaluated
according to the formalism presented in Sect.~\ref{sec3} using the
deuteron wave function for the Bonn OBEPQ potential of~\cite{MaH87}. 
For the final state interaction effects we have included in $KN$
scattering partial waves up to $\ell=1$, in $YN$ scattering up 
to $j=1$, and in $\pi N\rightarrow KY$ only $\ell=0$. 

\begin{figure}[htbp]
\includegraphics[width=.6\textwidth]{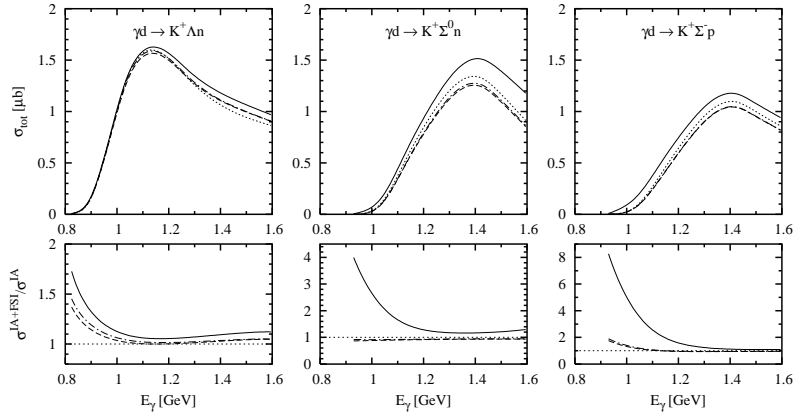}
\caption{Total cross sections of kaon photoproduction on
the deuteron (top) and their ratios to the impulse
approximation (bottom) for the separate channels $\gamma d\rightarrow
K^+\Lambda n$ (left), $\gamma d\rightarrow K^+\Sigma^0 n$
(middle), and $\gamma d\rightarrow K^+\Sigma^- p$ (right). Notation of
curves: dotted: impulse approximation (IA); dash-dot:
IA+YN rescattering; dashed: IA+YN+KN rescattering; solid: 
IA+YN+KN+$\pi N \rightarrow K Y$.} 
\label{fig-resdisc-gdkyn_tc}
\end{figure}

We will begin with a discussion of the total cross sections for the
three possible $K^+$-channels as shown in
Fig.~\ref{fig-resdisc-gdkyn_tc}. In the upper panels we show the various 
effects starting with the pure impulse approximation and then adding
successively $YN$ rescattering, $KN$ rescattering and finally the pion 
mediated contribution. In order to give a more detailed and 
quantitative evaluation we show
in the lower panels of Fig.~\ref{fig-resdisc-gdkyn_tc} the relative
effects by plotting the ratios of the corresponding cross sections to
the ones for the IA. 

\begin{figure}[htbp]
\includegraphics[width=.6\textwidth]{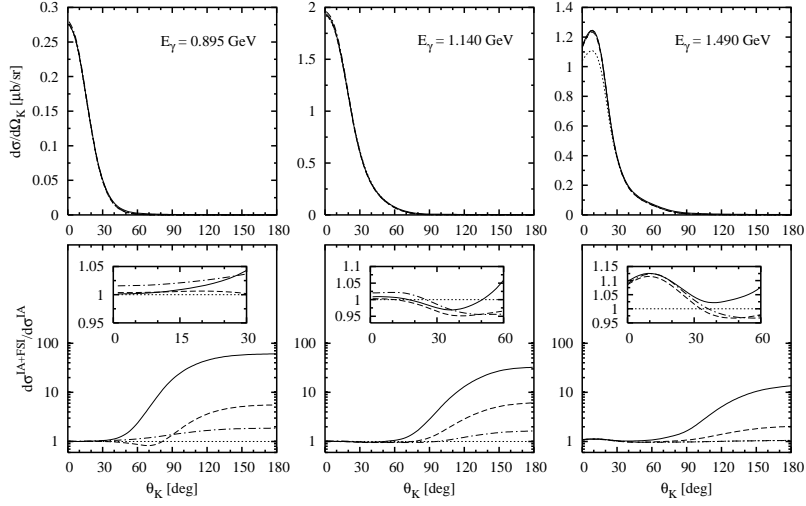}
\caption{Semi-inclusive differential cross section $d\sigma/d\Omega_K$ 
for $\gamma d\rightarrow K^+\Lambda n$ for different photon lab
energies with various interaction effects (top panels) and their
ratios with respect to the impulse approximation on a logarithmic
scale (bottom panels). The insets show the ratios at forward angles on
a larger linear scale. Notation of curves as in
Fig.~\ref{fig-resdisc-gdkyn_tc}.}  
\label{fig-resdisc-gdkyn1_dc}
\end{figure}

\begin{figure}[htbp]
\includegraphics[width=.6\textwidth]{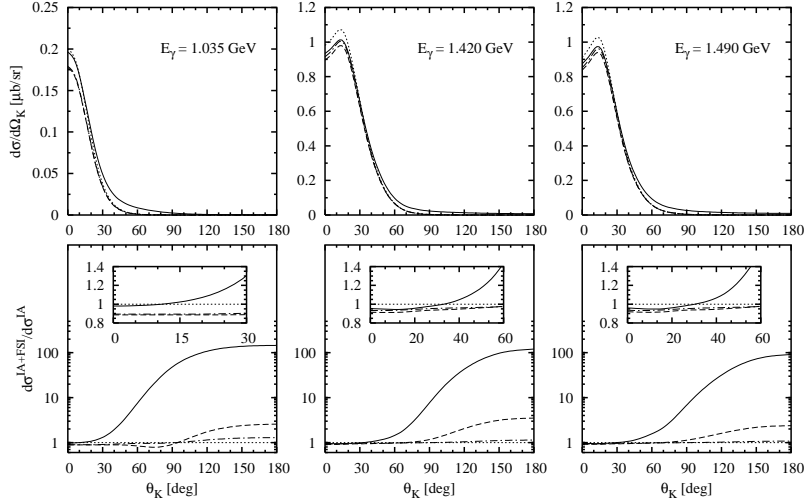}
\caption{Semi-inclusive differential cross section $d\sigma/d\Omega_K$
for $\gamma d\rightarrow K^+\Sigma^{0} n$ for different photon lab
energies with various interaction effects (top panels) and their
ratios with respect to the impulse approximation on a logarithmic
scale (bottom panels). The insets show the ratios at forward angles on
a larger linear scale. Notation of curves as in
Fig.~\ref{fig-resdisc-gdkyn_tc}.}  
\label{fig-resdisc-gdkyn3_dc}
\end{figure}

One readily notes that $KN$ rescattering -- 
the difference between the dash-dot and the dashed curves -- is
quite small, almost completely negligible for the total cross section 
for all three channels. 
For $\gamma$d $\rightarrow K^+ \Lambda n$ (left panel)
one notes a sizeable enhancement near the threshold, which
originates from $YN$ rescattering and the pion mediated reaction, 
comparable in size. At higher photon energies the enhancement is reduced
to about 10~\% with a slight dominance of the pion mediated process.
The reaction $\gamma d\rightarrow K ^+ \Sigma^0 n$ (middle panel) 
shows a different behaviour. While $YN$ rescattering decreases 
the cross section slightly by about $5-8$~\% over the whole range of 
photon energies, the pion mediated contribution acts in the opposite direction
leading to an overall increase compared to the IA, which is
quite dramatic near threshold and is still about $20$~\% at
higher energies. Finally, one finds for the channel 
$\gamma d\rightarrow K ^+ \Sigma^-p$ (right panel) a significant
increase by $YN$ rescattering close to threshold but above 1~GeV a
small reduction of the cross section by about 5~\%, a tiny increase from
$KN$ rescattering, and as most dominant effect again the pion mediated 
process with a strong near-threshold increase which levels off above
1~GeV to about $10$~\% relative to the IA. 
The reason for the different influence of the pion mediated reaction 
lies in the fact that for 
$K^+\Lambda n$ only the isospin $t=1/2$ contributes to the process
$\pi N\rightarrow K\Lambda$, whereas for $\pi N\rightarrow K\Sigma$
also the more important $t=3/2$-contribution appears. 

\begin{figure}[htbp]
\includegraphics[width=.6\textwidth]{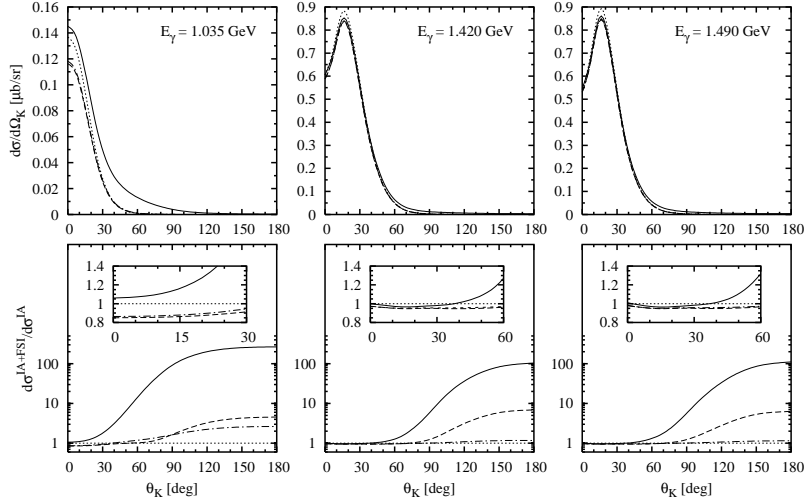}
\caption{Semi-inclusive differential cross section $d\sigma/d\Omega_K$
for $\gamma d\rightarrow K^+\Sigma^{-} p$ for different photon lab
energies with various interaction effects (top panels) and their
ratios with respect to the impulse approximation on a logarithmic
scale (bottom panels). The insets show the ratios at forward angles on
a larger linear scale. Notation of curves as in
Fig.~\ref{fig-resdisc-gdkyn_tc}.} 
\label{fig-resdisc-gdkyn5_dc}
\end{figure}

%%%%%%%%%%%%%%%%%%%%%%%%---DIFFERENTIAL---%%%%%%%%%%%%%%%%%%%%%%%%%%%%%%%%%%%%%

\begin{figure}[htbp]
\includegraphics[width=.4\textwidth]{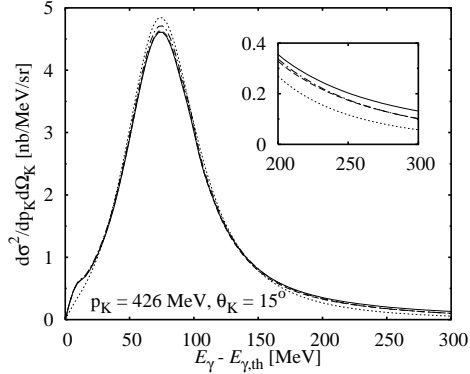}
\caption{Double differential cross section $d^2\sigma/dp_Kd\Omega_K$ 
for $\gamma d\rightarrow K^+\Lambda n$ for fixed $p_K$ and $\theta_K$ 
as function of photon excess energies above threshold 
with various interaction effects. Notation of curves as in
Fig.~\ref{fig-resdisc-gdkyn_tc}.}  
\label{fig_diff_cross_ex_new}
\end{figure}
As next topic we will discuss the semi-inclusive differential cross
section $d\sigma/d\Omega_K$ as defined in~(\ref{eq-gdkyn-dsigma/domegak}) 
for the three 
channels. The top panels of Fig.~\ref{fig-resdisc-gdkyn1_dc} show the
differential cross sections for the channel $\gamma d\rightarrow K ^+
\Lambda n$ at three photon lab-energies, one close to the threshold,
the next in the maximum of the total cross section, and finally the
third above the maximum. As one notes, the kaon is
produced predominantly into the forward direction, the maximum being at 
$0^\circ$ near threshold but moving to about $15^\circ$ at higher 
energies, while at 
backward angles the cross section drops rapidly. The reason for 
this behaviour is that for backward kaon production in the 
impulse approximation the spectator 
nucleon is forced to have a large momentum, for which, in turn, 
the deuteron wave function is strongly suppressed. In other words,
the probability for finding a spectator nucleon with the appropriate
momentum becomes increasingly tiny with increasing momentum transfer. 
In this situation 
it is more advantagous to share the large momentum transfer between 
both baryons as is provided by any of the two-step processes discussed 
here, particularly strong for the pion mediated process. 

\begin{figure}[htbp]
\includegraphics[width=.6\textwidth]{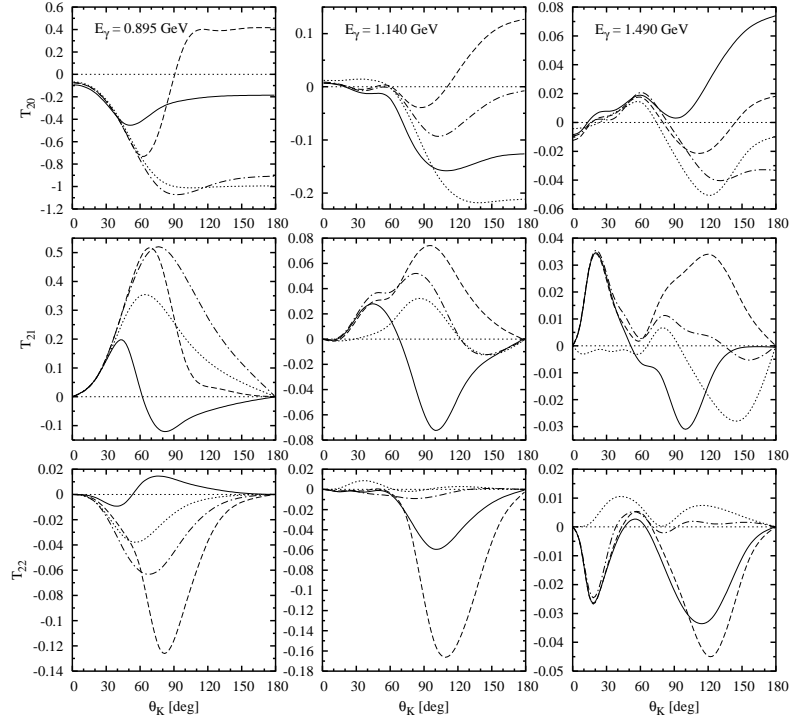}
\caption{Tensor target asymmetries for channel $\gamma d\rightarrow K
^{+} \Lambda n$ at different photon lab energies. Panels in the same
column refer to the same photon energy. Notation of curves as in
Fig.~\ref{fig-resdisc-gdkyn_tc}.}
\label{fig-resdisc-gdkyn1_tensor}
\end{figure}
In order to see more clearly the relative size of the interaction effects 
we have plotted in the lower panels of
Fig.~\ref{fig-resdisc-gdkyn1_dc} the ratios with respect to the IA.
The insets show the ratios for forward angles on a magnified linear scale.
In the forward direction $YN$ and $KN$ rescattering show 
at the two lower energies a small influence but opposite in sign so that 
the net effect is tiny. At the highest energy, one notes a sizeable 
increase at forward angles by $YN$ rescattering only of about 10~\%. 
In the backward direction the situation changes completely. 
The aforementioned mechanism of 
redistributing the large momentum transfer onto two
particles by two-step processes is most evident at backward angles, 
where one finds a huge increase of the cross section. Here 
$KN$ rescattering becomes more important than $YN$ rescattering, in 
particular at the highest energy, but most dominant is the pion 
mediated contribution. One also notes that the relative effect of the latter 
decreases with increasing energies.

\begin{figure}[htbp]
\includegraphics[width=.6\textwidth]{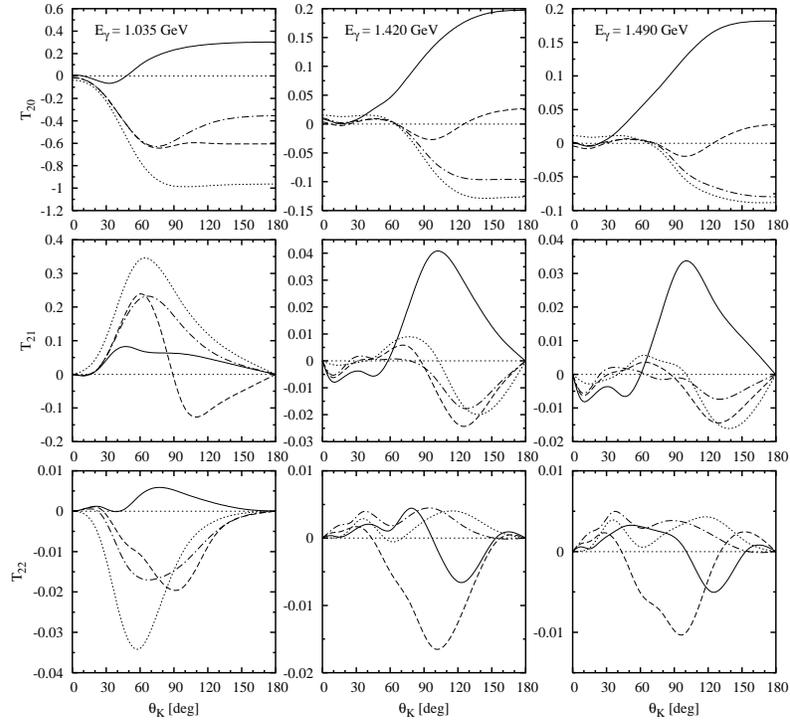}
\caption{Tensor target asymmetries for channel $\gamma d\rightarrow K
^{+} \Sigma^{0} n$ at different photon lab energies. Panels in the same
column refer to the same photon energy. Notation of curves as in
Fig.~\ref{fig-resdisc-gdkyn_tc}.}
\label{fig-resdisc-gdkyn3_tensor}
\end{figure}

\begin{figure}[htbp]
\includegraphics[width=.6\textwidth]{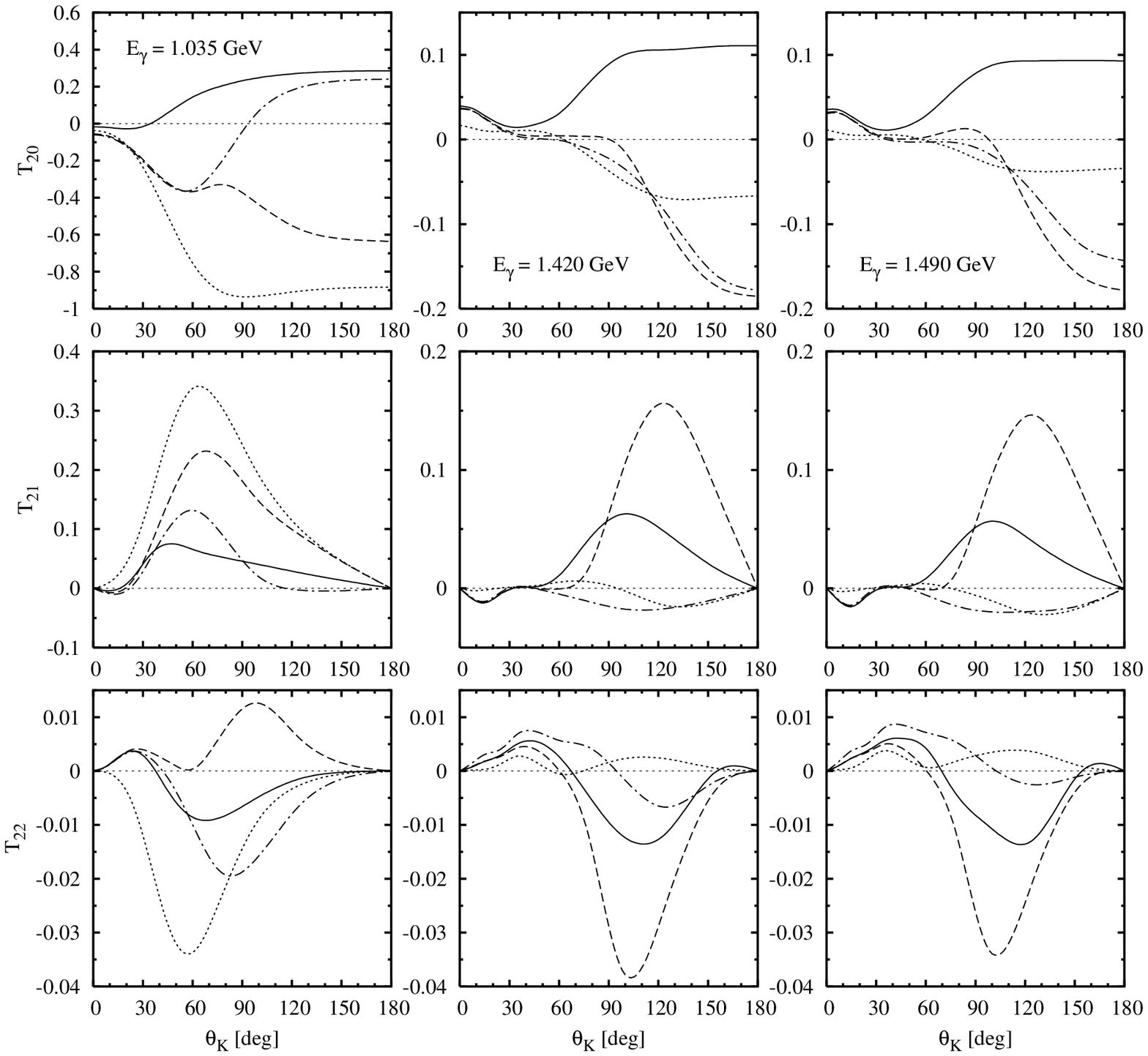}
\caption{Tensor target asymmetries for channel $\gamma d\rightarrow K
^{+} \Sigma^{-} p$ at different photon lab energies. Panels in the same
column refer to the same photon energy. Notation of curves as in
Fig.~\ref{fig-resdisc-gdkyn_tc}.}
\label{fig-resdisc-gdkyn5_tensor}
\end{figure}

The analogous results for the channel $\gamma d\rightarrow K^+\Sigma^{0}n$ 
are shown in Fig.~\ref{fig-resdisc-gdkyn3_dc}. Contrary to the foregoing 
case, one notes for this channel in the forward direction 
at all photon energies a sizeable reduction from $YN$ rescattering 
which, however, is partially counterbalanced at the lowest energy by the 
$\pi\rightarrow K$ process, while $KN$ rescattering is negligible. 
Again the pion process becomes dominant at backward angles
(see lower panels). Furthermore, $KN$ rescattering shows there
a small effect like for $\gamma d\rightarrow K ^+\Lambda n$, and
$YN$ rescattering is marginal here. 
For the other $\Sigma$-channel, $\gamma d\rightarrow K^+ \Sigma^{-}p$, 
the differential cross sections in Fig.~\ref{fig-resdisc-gdkyn5_dc}
display qualitatively the same features, the reduction of the cross 
section in forward direction by $\Sigma N$ rescattering, which is more than 
compensated by the $\pi\rightarrow K$-process at the lowest energy and 
partially diminished at the other two energies, and a very strong 
enhancement at backward angles. The influence of $KN$ rescattering 
remains relatively small.

For a comparison with previous results of Refs.~\cite{ReR67b} 
and \cite{Ker01} we have evaluated for the channel 
$\gamma d\rightarrow K ^+\Lambda n$ 
the double differential cross section $d^2\sigma/dp_K d\Omega_K$ for
the same kinematics, i.e.\ fixed kaon momentum $p_K=426$~MeV and kaon angle 
$\theta_K=15^\circ$. The result is shown in Fig.~\ref{fig_diff_cross_ex_new} 
as function of the excess photon energy above threshold. The enhancement
by $YN$ rescattering close to threshold is very similar to \cite{ReR67b,Ker01}.
In this energy region the $\pi\rightarrow K$ process shows little effect. 
However, above the peak this process becomes increasingly more important 
as one readily notes in Fig.~\ref{fig_diff_cross_ex_new}. A 
comparison to the work of Maxwell~\cite{Max04} is not possible since
in~\cite{Max04} total and semi-inclusive cross sections were not given. 

As last topic, we will discuss the influence of interaction
effects on polarization observables for which we have chosen the tensor 
target asymmetries for a tensor polarized deuteron target. 
Fig.~\ref{fig-resdisc-gdkyn1_tensor} shows the three types of asymmetries
$T_{20}$ (top panels), $T_{21}$ (middle panels), and $T_{22}$ (bottom panels)
for the channel $\gamma d\rightarrow K ^{+} \Lambda n$ at the same 
photon lab energies as for the differential cross sections. As can be
seen in the top panels of Fig.~\ref{fig-resdisc-gdkyn1_tensor}, the 
tensor target asymmetry $T_{20}$ is relatively small at 
forward angles for all given photon energies, and interaction effects 
show little influence, while at backward angles they become much more
pronounced. Their effect changes quite strongly with energy. At the
lowest energy $YN$ rescattering shows little influence whereas $KN$
rescattering changes $T_{20}$ from $-1$ for IA to 0.4 which is then
reduced to $-0.2$ by the pion process. At $E_\gamma=1.14$~GeV $YN$
rescattering reduces $T_{20}$ at 180$^\circ$ from $-0.2$ for IA to almost
zero, then it is increased to 0.13 by $KN$ rescattering to be reduced again
to $-0.12$ by the pion contribution. For the highest energy $T_{20}$
is changed from $-0.01$ for IA to 0.07 by all interaction effects. 
 
For $T_{21}$ and $T_{22}$ in the middle and bottom panels of
Fig. \ref{fig-resdisc-gdkyn1_tensor}, respectively, one also notes
quite a variation with the various interaction contributions, their
relative influence varying quite strongly with photon energy. The 
$\pi\rightarrow K$ process counteracts in general the influence of
$YN$ and $KN$ rescattering. The final results for $T_{21}$ show a
positive forward peak with a size between 0.2 at the lowest energy and
0.03 at the highest one, and a negative minimum around 90$^\circ$ of
$-0.1$ to $-0.03$. $T_{22}$ develops a negative minimum around
100-110$^\circ$ of the order of 0.06-0.03 for the highest two energies.

The tensor target asymmetries for channel 
$\gamma d\rightarrow K ^{+} \Sigma^{0} n$ are shown in
Fig.~\ref{fig-resdisc-gdkyn3_tensor}. The top panels of this figure
show that the values of $T_{20}$ are again relatively small in the forward
region. The various rescattering contributions add constructively so
that at backward angles the negative values for $T_{20}$ in IA is
changed to positive values of 0.3 to 0.18 going from the lowest to the
highest energy. At the highest energy, $YN$ rescattering has quite a
small effect as shown in the right top panel of
Fig.~\ref{fig-resdisc-gdkyn3_tensor}. In the middle panels one
can see that $KN$ rescattering has quite a strong effect on $T_{21}$
at the lowest energy but becomes small for the two higher energies
where the two-step process results in a positive maximum of about 0.04.
Remarkable influence of $KN$ rescattering is seen in 
$T_{22}$ in the bottom panels of
Fig.~\ref{fig-resdisc-gdkyn3_tensor}. But the two-step process
counteracts again so that altogether quite a small size of the order
0.005 results. 

Finally, Fig.~\ref{fig-resdisc-gdkyn5_tensor} exhibits the tensor target
asymmetries for the reaction $\gamma d\rightarrow K
^{+}\Sigma^{-}p$. For the asymmetry $T_{20}$ one readily notes a
sizeable influence from $YN$ rescattering while $KN$ rescattering
shows a strong effect only for the lowest energy. Again the pion
process acts in the opposite direction resulting in a total value of
0.2 to 0.1 at backward angles. In $T_{21}$ $KN$ rescattering exhibits
a remarkable influence which, however, is compensated by the pion
contribution. A similar situation is found for $T_{22}$. The total
result shows a small forward positive peak and a negative minimum at
higher angles of the order $-0.01$. 

%%%%%%%%%%%%%%%%%%%%%%%%  summary   %%%%%%%%%%%%%%%%%%%%%%%%%%%%%%%%%%%

\section{Summary and conclusions}
\label{sec5}

Kaon photoproduction on the deuteron has been investigated with respect 
to the importance of final state interactions, i.e.\ $YN$  and $KN$ 
rescattering, and of a two-body photoproduction contribution in terms of 
a pion mediated process 
$\gamma d \rightarrow \pi NN \rightarrow KYN$. The latter turned out
to give the dominant contribution beyond the IA, except for 
$\gamma d\rightarrow K^{+}\Lambda n$ near threshold where the influence 
of $\Lambda N$ rescattering is comparable to the pion process. The latter
leads to an increase of the total cross section relative to the IA. The 
increase is particularly large in the near threshold region for the 
$K\Sigma$ channels. This effect has its origin in the considerably 
stronger pion photoproduction amplitude on the nucleon compared to 
the corresponding amplitude for the kaon photoproduction in combination with
a sizeable $\pi N\rightarrow KY$ strangeness exchange reaction. 
Next in importance is $YN$ rescattering while $KN$ rescattering is much 
smaller.
The overall enhancement of the total cross section from all interaction 
effects is about 2~\% at the peak for 
$\gamma d\rightarrow K ^+\Lambda n$, 10~\%
for $\gamma d\rightarrow K ^+\Sigma^0 n$, and 7~\% for 
$\gamma d\rightarrow K ^+\Sigma^-p$. 

The semi-inclusive differential cross sections show that the kaon is 
mostly produced in the forward direction where the impulse approximation 
works reasonably well. But for a precise description, at least the effects 
from YN rescattering and the $\pi\rightarrow K$-process have to be considered. 
At backward angle, the strongest enhancement arises from the pion mediated 
process. As expected on general grounds, the tensor target asymmetries 
$T_{20}$, $T_{21}$, and $T_{22}$ exhibit a much stronger sensitivity 
to the various interaction effects, in particular at backward angles. 

Therefore, we may state as general conclusion 
that for studying the elementary kaon photoproduction on the neutron in 
the reaction on the deuteron the influence of interaction effects 
have to be cleanly separated using a reliable theoretical model. The 
same caveat applies for the study of the hyperon-nucleon interaction 
in this reaction. 

At present such interaction effects are treated in an approximate way by 
including them completely only in the two-body subsystems. Thus an
extension to a three-body formalism is desirable for the future in 
kinematic regions where 
such interaction effects appear substantial. Furthermore, instead of 
the simple separable potentials used in the present work, more realistic 
potentials for $KN$ scattering and the $\pi\rightarrow K$ process should 
be considered. Also the model for the electromgnetic production operator
could be improved, in particular the question of gauge invariance should
be addressed more carefully. Moreover, the present formalism can
also be extended to study kaon electroproduction on the deuteron 
in order to exploit the additional degrees of freedom of virtual photons.

\section*{Acknowledgements}
We would like to thank M.\ Schwamb and A.\ Fix for useful
discussions and a critical reading of the manuscript. Special thanks
go to K.\ Miyagawa for allowing us to use his $YN$ rescattering code. 
A.\ Salam acknowledges a fellowship from Deutscher 
Akademischer Austauschdienst (DAAD) and would like to thank the
Institut f\"ur Kernphysik of the Johannes Gutenberg-Universit\"at, 
Mainz for the very kind hospitality.

\end{document}